\documentclass[twoside,11pt]{article}
\usepackage{a4}
\usepackage{bm}
\usepackage{amsmath,bbm,fancyhdr,epsfig,amssymb,psfrag,multirow,exscale,caption,epsfig,slashbox,color}
 \allowdisplaybreaks[1] 
\numberwithin{equation}{section}
\newcommand{\be}{\begin{equation}}
\newcommand{\ee}{\end{equation}}
\newcommand{\bea}{\begin{eqnarray}}
\newcommand{\eea}{\end{eqnarray}}
\newcommand{\ba}{\begin{align}}
\newcommand{\ea}{\end{align}}
\newcommand{\refe}[1]{Eq.\ (\ref{#1})}
\newcommand{\refes}[2]{Eqs.\ (\ref{#1}--\ref{#2})}
\newcommand{\refec}[2]{Eqs.\ (\ref{#1},\:\ref{#2})}

\newcommand{\R}{\mathbbm{R}}

\newcommand{\uosp}{U\!O\!Sp(1|2)}

\newcommand{\extd}{\mathrm{d}}
\newcommand{\extcov}{\mathrm{D}}
\newcommand{\dd}[1]{\frac{\mathrm{d}}{\mathrm{d}#1}}
\newcommand{\defined}{\stackrel{\mathrm{def}}{=}}
\newcommand{\ie}{i.e.\ }
\definecolor{grey}{rgb}{0.45,0.45,0.45}

\title{
\vskip -70pt
\begin{flushright}
{\normalsize \ DAMTP-2006-83}\\
\end{flushright}
\vskip 25pt {\bf Super coset space geometry} } \vspace{1.4cm}
{\makeatletter
\author{A.~F.~Kleppe\thanks{e-mail address: A.F.Kleppe@damtp.cam.ac.uk}
\thanks{AFK previously published under the name A.\ F.\
Schunck.} \hspace{1pt} and Chris Wainwright{\thanks{e-mail address:
C.J.Wainwright@damtp.cam.ac.uk}} \\ \small{\textsl{Department of Applied
    Mathematics and Theoretical Physics}}
\\ \small{\textsl{University of Cambridge}} \\
\small{\textsl{Wilberforce Road, Cambridge CB3 0WA, England}}}
\makeatother}
\date{\today}
\begin{document}
\maketitle

\begin{abstract}
Super coset spaces play an important role in the formulation of
supersymmetric theories. The aim of this paper is to review and discuss
the geometry of super coset spaces with particular focus on the way the
geometrical structures of the super coset space $G/H$ are inherited from
the super Lie group $G$. The isometries of the super coset space are
discussed and a definition of Killing supervectors -- the supervectors
associated with infinitesimal isometries -- is given that can be easily
extended to spaces other than coset spaces.

\vspace{3ex} \noindent PACS numbers: 02.40.-k, 11.30.Pb, 12.60.Jv

\end{abstract}

\section{Introduction}
\label{sec:intro} Coset spaces are widely used in a variety of contexts,
for example as target spaces within string theory or more generally within
non-linear sigma models. For supersymmetric theories of this kind, a
thorough understanding of the geometry of super coset spaces is therefore
essential. On the other hand, supersymmetric theories with coset spaces as
the base naturally have a superspace formulation in terms of super coset
spaces, the most prominent examples being supersymmetric theories in flat
space. These can be formulated in terms of flat superspace, which is the
quotient of the super Poincar\'{e} group by its Lorentz subgroup. We aim
to provide in this paper a firm mathematical basis for the geometry of
super coset spaces, collating and making rigorous results often only
sketched in the literature and extending results only discussed for
ordinary coset spaces to the supersymmetric case.

The geometry of a (super) coset space $G/H$, i.e.\ its frame and
connection, can be determined in terms of the geometry of $G$. In fact, it
is well known that by pulling back the Maurer-Cartan one-form on the group
$G$ to the base $G/H$ one can obtain the frame and connection on the base,
see e.g.\ \cite{coset,hep-th/0503196}. While this is often stated in the
literature a geometric explanation of this is usually omitted. Treating
the super Lie group $G$ as a principal bundle over $G/H$ we give a
self-contained account of the geometry of coset spaces focussing in
particular on the relation between the geometry in the bundle and the
geometry in the base.

Coset spaces are characterized by high symmetry; most of their isometries
can be derived from the left action of $G$ on the coset space and as such
the isometry group of $G/H$ contains $G$. On the other hand, the
isometries of the coset space can be determined directly from its
geometry: In the case of ordinary space the infinitesimal isometries,
i.e.\ the Killing vector fields, are defined as the directions along which
the metric tensor is dragged into itself. In the case of superspace --
where the tangent space group of the supermanifold under consideration
must correspond to the even Grassmann extension of the tangent space group
of the body of the supermanifold, i.e.\ to $SO(p,q)$ -- such a definition
is no longer feasible. This is as, in this case, there exists no
physically natural superspace generalization of the concept of metric
\cite{bluebook,west}: If one were to introduce a supermetric the tangent
space group of the supermanifold derived from this would be too large, and
hence would not correspond to the even Grassmann extension of $SO(p,q)$,
thus rendering the theory unphysical. While this problem is often noted in
the literature its relevance to the definition of Killing supervectors is
seldom elucidated. One aim of this paper is therefore to make rigorous and
clarify the notion of Killing supervector fields in the context of
superspace. As we shall see it is possible to define Killing supervector
fields as those infinitesimal transformations that leave the frame and
connection invariant up to a gauge transformation, see also
\cite{bluebook}. This condition can be rephrased in terms of the
commutator of some generalized Lie derivative and the covariant
derivative, cf.\ \cite{bluebook}. In the case of super coset spaces we
shall see that this definition reproduces the Killing supervectors derived
from the left action of the group on the super coset space which justifies
this definition also for more general spaces.

Derivations such as the covariant derivative and generalized Lie
derivative are an integral part of the geometry of coset spaces. We give a
geometric interpretation to derivations such as these on $G/H$ by defining
them in terms of maps on the bundle $G$ and associated bundles.

The organization of this paper is as follows: First we briefly review the
construction of $G$ as a principal bundle over the coset space $G/H$. We
then discuss the geometry of $G$ focussing in particular on its invariant
vector fields and the Maurer-Cartan form in the bundle. In Sections
\ref{sec:connection}--\ref{sec:curvature} we discuss in detail how the
geometry in the base, i.e.\ connection and frame, torsion and curvature,
can be obtained from the geometry in the bundle. In Section
\ref{sec:bundle} we introduce associated bundles and briefly review some
aspects of supertensor bundles that will be important when discussing
derivations on associated bundles. We then introduce those (local) bundle
maps that will allow us to define in Section \ref{sec:derivations}
derivations on supertensor bundles, such as the covariant derivative, the
Lie derivative and the so-called $H$-covariant Lie derivative. As we shall
see in Section \ref{sec:killing}, Killing supervectors, i.e.\ the
supervectors associated to infinitesimal isometries, can be defined in
terms of a so-called generalized Lie derivative which combines an
arbitrary transformation in the base with an arbitrary gauge
transformation in the bundle, and which we require to commute with the
covariant derivative. Finally, in Section \ref{sec:example}, we shall
apply the concepts and methods introduced in the previous sections to flat
superspace as an example.

\section{$G$ as a principal bundle over $G/H$}\label{sec:principal}
Consider a super Lie group $G$ with super Lie subgroup $H$. A super Lie
group shall be defined in the sense of DeWitt, see \cite{dewitt}, as a
group which is also a supermanifold and which has a differentiable group
multiplication. We define the super coset space $G/H$ via the equivalence
of group elements in $G$ under right multiplication by an element of $H$,
\begin{equation}
G/H=\{gH:g\in G\}.
\end{equation}
The coset space $G/H$ naturally inherits a supermanifold structure from
the supermanifold $G$ \cite{dewitt}. The geometry of the coset space $G/H$
is also inherited from $G$. To study this we consider $G$ as a principal
bundle over $G/H$ with fibre $H$. The construction in the
non-supersymmetric case \cite{nakahara, steenrod} is directly transferred
to the supersymmetric case. First we have the bundle projection map
\begin{align}
\pi &: G \to G/H\nonumber \\
&: g \mapsto gH.
\end{align}
The inverse image $\pi^{-1}(p)$ of a point $p$ in the base gives us the
fibre above that point which is clearly isomorphic to $H$. To define the
local trivializations of the bundle we consider charts $U_i\subset G/H$ on
the base. Within a particular chart it is always possible to choose a
local section. By a local section we mean a map $L_i:U_i\to \pi^{-1}(U_i)$
which satisfies $\pi\circ L_i=\mathrm{id}_{U_i}$. The local section $L_i$
provides us with a coset representative $L_i(p)$ for any $p \in G/H$.
Using this local section we define the canonical local trivialization for
the bundle
\begin{align}
\phi_i&:U_i\times H \to \pi^{-1}(U_i)\nonumber\\
\label{eqn:localtriv}&:(p,h)\mapsto L_i(p)h.
\end{align}
Note that the inverse is easily constructed as $\phi_i^{-1}(g)=(\pi(g),
L_i(\pi(g))^{-1}g)$. The transition functions for $U_i\cap U_j\ne
\mbox{\Large$\varnothing$}$ are defined as
\begin{equation}
t_{ij}(p)=\phi_{i,p}^{-1}\phi_{j,p}:H\to H,
\end{equation}
where $\phi_{i,p}(h)\equiv \phi_i(p,h)$. The map $t_{ij}(p)$ is clearly
just left multiplication by an element of $H$, and we shall use the
notation $t_{ij}(p)(h)=t_{ij}(p)h$. The trivializations are then related
as
\begin{equation}
\phi_j(p,h)=\phi_i(p,t_{ij}(p)h).
\end{equation}
The structure group is therefore $H$ and acts on the fibre $H$ by left
multiplication. We thus have that $G$ is a principal bundle over $G/H$
with structure group $H$. We denote this as $G=P(G/H,H)$.

\section{Group geometry}
\subsection{Invariant vector fields}\label{sec:invvecfields}
Let us denote the super Lie algebra associated to the super Lie group $G$
by $\mathfrak{g}$. The generators of $\mathfrak{g}$ will be denoted by
$T_p$, $p=1,\ldots,\mathrm{dim}\,\mathfrak{g}$; they have definite parity,
either even or odd and we will set $(-1)^p = 1$ for $T_p$ even and $(-1)^p
= -1$ for $T_p$ odd. The index in the exponent of $(-1)$ is as such to be
understood as taking the values $0$ or $1$ according to whether it is even
or odd. A general element of the super Lie algebra is then expanded in the
generators as $X = X^p T_p$, where the $X^p$ are pure supernumbers chosen
such that $X$ is even. As such the super Lie algebra consists of even
elements only, in fact $\mathfrak{g}$ can be viewed as the even part of a
larger Berezin superalgebra \cite{bluebook}. The super Lie group $G$ can
then be obtained from its super Lie algebra via the exponential mapping.

For each element of the algebra $A\in\mathfrak{g}$ we can construct two
different supervector fields, $A^\sharp$ and $A^\flat$, defined by their
action on a function $f$ on $G$ as
\begin{subequations}
\begin{align}
\label{eqn:sharp}A^\sharp\big|_g[f]&\defined\dd{t}\left(f\left(ge^{tA}\right)\right)\Big|_{t=0}\\
\label{eqn:flat}A^\flat\big|_g[f]&\defined\dd{t}\left(f\left(e^{tA}g\right)\right)\Big|_{t=0}.
\end{align}
\end{subequations}
We have thus obtained two maps from the algebra $\mathfrak{g}$ to the
space of vector fields on $G$ given by $\sharp:A\mapsto A^\sharp$ and
$\flat:A\mapsto A^\flat$. Clearly these maps are linear.

The definition of the vector fields given above may be modified in the
case that $A$ is an odd element of the Berezin superalgebra simply by
choosing the parameter $t$ to be an odd supernumber. This ensures that
$e^{tA}$ is still a group element. This way it possible to define the
supervectors $T_p{^\sharp}$ and $T_p{}^\flat$ for all $p$.

It is easy to see that $A^\sharp$ is a left-invariant vector field,
whereas $A^\flat$ is right-invariant:
\begin{subequations}
\begin{align}
{L_{g_1}}_\ast \left(A^\sharp\big|_g\right)&=A^\sharp\big|_{g_1g}\\
{R_{g_1}}_\ast \left(A^\flat\big|_g\right)&=A^\flat\big|_{gg_1}.
\end{align}
\end{subequations}
Here $L_g$ and $R_g$ are the group operations of left and right
multiplication by $g\in G$, the lowered asterisk ($*$) is used to denote
the corresponding induced map on vector fields (the pushforward). The
vector fields also satisfy
\begin{subequations}
\begin{align}
\label{eqn:rightactiononsharp}{R_{g_1}}_\ast \left(A^\sharp\big|_g\right)
&=
\left(\mathrm{Ad}_{{g_1}^{-1}}A\right)^\sharp \big|_{gg_1}\\
{L_{g_1}}_\ast \left(A^\flat\big|_g\right) &=
\left(\mathrm{Ad}_{g_1}A\right)^\flat \big|_{g_1g}.
\end{align}
\end{subequations}
Here $\mathrm{Ad}_g$ is the adjoint action of the group on its algebra
which is induced from the adjoint action of the group on itself. For the
latter we also use the notation $\mathrm{Ad}_g$. We have
$\mathrm{Ad}_gh=ghg^{-1}$ and thus, in a matrix representation, the
adjoint action on the algebra is thus just $\mathrm{Ad}_gA=gAg^{-1}$.

Under the Lie bracket of supervector fields we find
\begin{subequations}
\begin{align}
[A^\sharp,B^\sharp]&=[A,B]^\sharp\\
\label{eqn:flatalgebra} [A^\flat,B^\flat]&=-[A,B]^\flat\\
[A^\sharp,B^\flat]&=0.
\end{align}
\end{subequations}
The bracket occurring on the right is the super Lie algebra bracket.

\subsection{The Maurer-Cartan one-form}\label{sec:maurercartan}
The Maurer-Cartan one-form $\zeta$ is a super Lie algebra valued one-form
defined on the super Lie group as
\begin{equation}\label{eqn:maurercartan}
\zeta(A^\sharp) \defined A, \quad \forall A\in\mathfrak{g}.
\end{equation}
Note that one should be careful when dealing with forms acting on vectors
in the case of supersymmetry, our conventions are given in Appendix
\ref{sec:formconventions}.

The Maurer-Cartan one-form is clearly left-invariant, whereas under right
translations it transforms as
\begin{equation}\label{eqn:rightactionmaurercartan}
R_g^*\zeta = \mathrm{Ad}_{g^{-1}} \zeta.
\end{equation}
The Maurer-Cartan form can be shown to satisfy
\begin{equation}\label{eqn:maurercartanstructure}
\extd\zeta+\zeta\wedge\zeta=0,
\end{equation}
which is the so-called \emph{Maurer-Cartan structure equation}.

When working in a matrix representation of the group the Maurer-Cartan
form may be represented as $\zeta=g^{-1}\extd g$
\cite{goeckelerschuecker}. Throughout the remainder of this paper we will
occasionally work in a matrix representation where it is more convenient.

\section{The connection on $G/H$}\label{sec:connection}
In this section we will construct a connection on the coset space $G/H$.
We will see that this connection is naturally induced from the group $G$,
in particular from the Maurer-Cartan form. A connection on a fibre bundle
is defined quite abstractly in terms of a decomposition of the tangent
space into so-called vertical and horizontal subspaces. While this
definition may not be the definition of connection familiar from physics
it will provide us with a deeper insight into the geometry. In Section
\ref{sec:localcon} we will review how this abstract definition relates to
the familiar notion of connection in physics.

\subsection{Horizontal and vertical subspaces}
Geometrically a connection on the bundle $G$ can be thought of as a
decomposition of the tangent space to the super Lie group,
$T_gG=T^{\mathrm{v}}_gG\oplus T^{\mathrm{h}}_gG$, into vertical and
horizontal subspaces, respectively. This decomposition must be supersmooth
and the horizontal subspaces must satisfy
\begin{equation}\label{eqn:horizontalrighttranslation}
T^{\mathrm{h}}_{gh}G={R_h}_\ast T^{\mathrm{h}}_gG.
\end{equation}
Such a decomposition of the tangent space can be achieved by utilizing the
left-invariant vector fields and the natural decomposition of the algebra
arising from the subgroup $H$.

Let us denote the super Lie algebra of the subgroup $H$ by $\mathfrak{h}$.
We then choose a subspace $\mathfrak{k}$ in $\mathfrak{g}$ complementary
to $\mathfrak{h}$, i.e.
\begin{equation}\label{eqn:algebrasplit}
\mathfrak{g}=\mathfrak{k}\oplus\mathfrak{h}.
\end{equation}
The generators of the full algebra $T_p$,
$p=1,\ldots,\mathrm{dim}\,{\mathfrak{g}}$, can then be split up into the
generators of $\mathfrak{h}$, $H_I$,
$I=1,\ldots\,\mathrm{dim}\,\mathfrak{h}$, and the remaining generators
$K_A$, $A=1,\ldots\,\mathrm{dim}\,{\mathfrak{k}}$. The structure constants
$f_{pq}{}^r$ of $\mathfrak{g}$ are then defined by
\begin{subequations}
\begin{align}
\label{eqn:algebra1} [H_I, H_J] & =  f_{IJ}{}^K H_K \\
\left[H_I, K_A\right] & =  f_{IA}{}^J H_J + f_{IA}{}^B K_B \\
\label{eqn:algebra3} \left[K_A, K_B\right] & = f_{AB}{}^J H_J + f_{AB}{}^C
K_C.
\end{align}
\end{subequations}
The bracket here is the graded Lie bracket which satisfies the symmetry
$[T_p,T_q]=-(-1)^{pq}[T_q,T_p]$.

If $\mathfrak{k}$ can be chosen such that the structure constants
$f_{IA}{}^J$ vanish,
i.e.~$[\mathfrak{h},\mathfrak{k}]\subseteq\mathfrak{k}$, then the group
$G$ is said to be \emph{reductive}. As we shall see this is an important
property and we shall assume that $G$ is reductive throughout the
remainder of this paper.

The decomposition of the algebra in \refe{eqn:algebrasplit} naturally
gives a decomposition of the tangent space to the group into horizontal
and vertical subspaces. We may use the generators of the algebra to define
a basis of the vertical and horizontal subspaces. We take
$\{H_I{}^\sharp\}$ to be a basis of $T^{\mathrm{v}}_gG$ and
$\{K_A{}^\sharp\}$ to be a basis of $T^{\mathrm{h}}_gG$, i.e.~
\begin{subequations}
\begin{align}
\label{eqn:vertical} T^{\mathrm{v}}_gG&\equiv\{X^IH_I{}^\sharp\big|_g :
X^I\in \R_\infty\}\\
\label{eqn:horizontal} T^{\mathrm{h}}_gG&\equiv\{X^AK_A{}^\sharp\big|_g :
X^A\in \R_\infty\},
\end{align}
\end{subequations}
where here $\R_\infty$ denotes the real supernumbers. This decomposition
is clearly smooth and of the form $T_gG=T^{\mathrm{v}}_gG\oplus
T^{\mathrm{h}}_gG$. From \refe{eqn:rightactiononsharp} and the reductive
property f the group it is easily seen that
\refe{eqn:horizontalrighttranslation} is satisfied.

Given a notion of horizontal and vertical vectors, it is natural to define
horizontal and vertical differential forms. A differential form on the
bundle is called vertical (respectively horizontal) if it vanishes
whenever one of the vectors on which it is evaluated is horizontal
(respectively vertical). For example, expanding the Maurer-Cartan form,
\refe{eqn:maurercartan}, in the algebra generators we have
\begin{equation}\label{eqn:maurercartandecomp}
\zeta=\zeta^AK_A+\zeta^IH_I.
\end{equation}
It is then easy to see that
\begin{alignat}{2}
\nonumber\zeta^A(K_B{}^\sharp) &= \delta_B{}^A &\qquad
\zeta^A(H_J{}^\sharp) &= 0 \\
\label{eqn:zetaKetc}\zeta^I(K_B{}^\sharp) &=0 & \zeta^I(H_J{}^\sharp) &=
\delta_J{}^I.
\end{alignat}
Hence $\zeta^A$ will vanish when acting on any vertical vector whereas
$\zeta ^I$ vanishes on any horizontal vector. $\zeta^A$ is therefore a
horizontal form and $\zeta^I$ a vertical form.
\refe{eqn:maurercartandecomp} can thus be thought of as a decomposition of
the Maurer-Cartan form into its horizontal and vertical parts.

\subsection{The connection one-form}
It is usually more convenient to define a connection in terms of a
\emph{connection one-form}. This is a super Lie algebra valued one-form,
$\Omega$, on the bundle and it is required to satisfy
\begin{subequations}
\begin{gather}
\label{eqn:connection1} \Omega(H_I{}^\sharp) =H_I, \quad
I=1,\ldots,\mathrm{dim}\,\mathfrak{h}\\
\label{eqn:connection2} R_h^\ast\Omega =\mathrm{Ad}_{h^{-1}}\Omega, \quad
\forall h \in H.
\end{gather}
\end{subequations}
The horizontal subspace is then defined as the kernel of $\Omega$, i.e.~
\begin{equation}
T^{\mathrm{h}}_gG=\{X\in T_gG : \Omega(X)=0\}.
\end{equation}

We will now show that the vertical part of the Maurer-Cartan form,
$\zeta^IH_I$, can be taken to be the connection one-form consistent with
the definition of horizontal and vertical subspaces defined in
\refec{eqn:connection1}{eqn:connection2}. Firstly, as $\zeta^AK_A$ is
horizontal, it follows immediately from \refe{eqn:maurercartan} that
\begin{subequations}
\begin{equation}
(\zeta^IH_I)(H_J{}^\sharp)=H_J, \quad
J=1,\ldots,\mathrm{dim}\,\mathfrak{h}.
\end{equation}
We also find immediately from \refe{eqn:rightactionmaurercartan} that
\begin{equation}
R_h^*(\zeta^IH_I) = \mathrm{Ad}_{h^{-1}} (\zeta^IH_I), \quad \forall h\in
H.
\end{equation}
\end{subequations}
These two equations can then be compared directly with
\refec{eqn:connection1}{eqn:connection2}. Also, as the form $\zeta^IH_I$
is vertical, its kernel is precisely the horizontal subspace. Thus we may
choose the connection one-form as $\Omega=\zeta^IH_I$.

\subsection{The local connection and parallel transport}\label{sec:localcon}
The definition of a connection in terms of horizontal and vertical
subspaces may seem a little abstract, and the connection one-form $\Omega$
is not the connection usually dealt with in physics. In this section we
shall review the relation of this abstract definition of connection to the
more familiar notion of connection in physics. To do this we shall
consider parallel transport.

Given a curve $\gamma:[0,1]\to G/H$ in the base we define a
\emph{horizontal lift} of $\gamma$ to be a curve $\tilde{\gamma}:[0,1]\to
G$ in the bundle which satisfies $\pi\circ\tilde{\gamma}=\gamma$ and for
which the tangent vector to $\tilde{\gamma}$ lies in
$T^{\mathrm{h}}_{\tilde{\gamma}(t)}G$. For each point
$g\in\pi^{-1}(\gamma(0))$ there is a unique horizontal lift of $\gamma$
for which $\tilde{\gamma}(0)=g$. Further, for $h \in H$, the horizontal
lift passing through $gh$ is simply $\tilde{\gamma}(t)h$.

Consider a point $g_0\in G$ above $\gamma(0)$, and construct the
horizontal lift $\tilde{\gamma}$ of $\gamma$ satisfying
$\tilde{\gamma}(0)=g_0$. Then the value $g_1=\tilde{\gamma}(1)$ is said to
be the \emph{parallel transport} of $g_0$ along $\gamma$.

Let us now analyze this construction as viewed from a local
trivialization. Let us choose the local section\footnote{We will drop
  subscript $i$'s where we only need to consider one chart.} such that
$L(p)=g_0$, we may then decompose the horizontal lift of $\gamma$ as
\begin{equation}\label{eqn:horizontallift}
\tilde{\gamma}(t)= L(\gamma(t))\tilde{h}(t),
\end{equation}
where $\tilde{h}(0)=1$. Then, if we let $X$ be the tangent vector to
$\gamma$ and let $\tilde{X}$ be the tangent vector to $\tilde{\gamma}$, it
is possible to show (see Appendix \ref{sec:appendix1}) that
\begin{equation}\label{eqn:Xmakehorizontal}
\tilde{X}={R_{\tilde{h}(t)}}_\ast(L_\ast X)+ (\tilde{h}(t)^{-1}\extd
\tilde{h} (X))^\sharp.
\end{equation}
Note that this result is here written in a matrix representation. Also,
$\extd$ is the exterior derivative on the base $G/H$ and should not be
confused with the exterior derivative on the group $G$ which we used in
Section \ref{sec:maurercartan}. Now, since $\tilde{X}$ is horizontal we
have that $\Omega(\tilde{X})=0$, hence using
\refec{eqn:connection1}{eqn:connection2} we find
\begin{align}
\nonumber 0&=\mathrm{Ad}_{\tilde{h}(t)^{-1}}\Omega(L_\ast X) +
\tilde{h}(t)^{-1}\extd
\tilde{h} (X) \\
\label{eqn:localconnectionmotivation} &= \tilde{h}(t)^{-1} \left(
L^\ast\Omega(X)\tilde{h}(t) + \dd{t}\tilde{h}(t) \right).
\end{align}
With this equation in mind we introduce the \emph{local connection},
$\omega^{(L)}$, a super Lie algebra valued one-form in $U$, defined as
\begin{equation}\label{eqn:localconnectiondefinition}
\omega^{(L)} \defined L^\ast\Omega.
\end{equation}
Thus we see from \refe{eqn:localconnectionmotivation} that
\begin{equation}\label{eqn:localconnectionproperty}
\dd{t}\tilde{h}(t) = -\omega^{(L)}(X)\tilde{h}(t).
\end{equation}
From this equation, or its formal solution in terms of the path ordered
exponential, we see that the local connection is precisely the connection
familiar to us from physics. This will become even more apparent in
Section \ref{sec:covder} when we work with associated bundles.

Note that $\omega^{(L)}$ clearly depends on the choice of local section.
The choice of local section can be seen to be precisely what we would
naturally call the choice of gauge. This view of gauge choice is one of
the many nice aspects of working in the language of bundles. To see this
in a little more detail let us consider the connection for two different
choices of local section: $L(p)$ and $L'(p)$. The two local sections can
clearly be related by a right multiplication, $L'(p)=L(p)h(p)$, with
$h(p)\in H$ dependent on $p\in U$. It is possible to show (see Appendix
\ref{sec:appendix1}) that for a vector $X\in T_pU$ we have
\begin{equation}\label{eqn:Xchangesection}
L'_\ast X = {R_{h}}_\ast(L_\ast X) + (h^{-1}\extd h (X))^\sharp.
\end{equation}
From this it follows directly that
\begin{displaymath}
\omega^{(L')}(X)=\omega^{(Lh)}(X)=h^{-1}\omega^{(L)}(X)h+h^{-1}\extd h(X).
\end{displaymath}
Since $X$ is arbitrary we deduce
\begin{equation}\label{eqn:gaugetransformconnection}
\omega^{(Lh)}=h^{-1}\omega^{(L)}h+h^{-1}\extd h,
\end{equation}
which is the usual transformation rule for a connection under the gauge
transformation given by the local field $h(p)\in H$.

\section{The frame and coframe on $G/H$}\label{sec:frame}
In the previous section we have discussed how the connection one-form on
the bundle $G$ can be taken to be the vertical part of the Maurer-Cartan
one-form. When pulled back under a local section this gives us the usual
connection on the base. It is then natural to ask what the horizontal part
of the Maurer-Cartan form is associated with. We will show in this section
that it is naturally associated with a frame.

Consider the pullback of the horizontal components of the Maurer-Cartan
form. We define
\begin{equation}\label{eqn:coframe}
E^A_{(L)}\defined L^\ast\zeta^A.
\end{equation}
Note that a coframe in $U$ is defined as a set of $\mathrm{dim}(G/H)$
one-forms that are linearly independent at each $p \in U$. Since $A = 1,
\ldots, \mathrm{dim} (G/H)$ there is clearly the right number of
$E^A_{(L)}$; all that remains is to show pointwise linear independence.
Consider a point $p \in U$. Suppose we introduce a set of supernumbers
$\lambda_A$ and impose that $\sum_AE^A_{(L)}|_p\lambda_A=0$. Acting on an
arbitrary vector $v\in T_pU$ and using the definition of pullback this
gives
\begin{displaymath}
\sum_A(\zeta^A\lambda_A)(L_\ast v)=0.
\end{displaymath}
Now, since $\zeta_A$ is horizontal it vanishes when acting on any vertical
vector $V_\mathrm{vert}\in T^{\mathrm{v}}_{L(p)}G$ and hence we have
\begin{displaymath}
\sum_A(\zeta^A\lambda_A)(L_\ast v+V_\mathrm{vert})=0.
\end{displaymath}
The vector $L_\ast v+V_\mathrm{vert}$ can be seen to be a completely
arbitrary supervector in $T_{L(p)}G$, this follows from the fact that an
arbitrary curve $g(t)\in\pi^{-1}(U)$ can be decomposed in terms of the
local trivialization as $g(t)=L(\gamma(t))h(t)$ where $\gamma(t)$ is a
curve in $U$, and $h(t)\in H$. So let us expand $L_\ast v+V_\mathrm{vert}$
in the basis of the $K_A^\sharp$ as $L_\ast
v+V_\mathrm{vert}=V^AK_A{}^\sharp$, where $V^A$ are therefore arbitrary
supernumbers of parity $A$ since $L_\ast v+V_\mathrm{vert}$ is even. We
then have
\begin{displaymath}
\sum_A V^A\lambda_A=0.
\end{displaymath}
Choosing $V^A$ appropriately we see that $\lambda_A=0$ for all $A$, hence
the $E^A_{(L)}$ are linearly independent.

The dual frame of vectors, $E_A^{(L)}$, satisfying
$E^B_{(L)}(E_A^{(L)})=\delta_A{}^B$, is given by
\begin{equation}\label{eqn:definecoframe}
E_A^{(L)}\big|_p=\pi_\ast \left(K_A{}^\sharp \big|_{L(p)}\right).
\end{equation}
This is easily checked as
\begin{align*}
E^B_{(L)}(E_A^{(L)}) &= L^\ast\zeta^B(\pi_\ast K_A{}^\sharp) \\
&= \zeta^B(L_\ast\pi_\ast K_A{}^\sharp)\\
&= \zeta^B(K_A{}^\sharp + V_{\mathrm{vert}}) \\
&= \delta_A{}^B.
\end{align*}
Here we have noted that the horizontal part of the supervector
$L_\ast\pi_\ast K_A{}^\sharp$ is just $K_A{}^\sharp$, its vertical part we
have called $V_{\mathrm{vert}}$ which vanishes when acted on by $\zeta^B$.
The final line then follows immediately from \refe{eqn:zetaKetc}.

Let us consider the behavior of the frame under a gauge transformation,
which, as remarked in Section \ref{sec:localcon}, is merely a change of
local section. We again introduce two local sections $L(p)$ and
$L'(p)=L(p)h(p)$ and will consider how the frame $E_A^{(L')}$ is related
to the frame $E_A^{(L)}$. First consider, for $A\in\mathfrak{k}$ and $f$
some arbitrary function on $G$,
\begin{align}
\pi_* \big(A^\sharp\big|_{L(p)h(p)} \big)[f] & =
A^\sharp\big|_{L(p)h(p)}[f \circ \pi]  \nonumber\\
& = \dd{t} \left(f \circ \pi \left(L(p)h(p)e^{t A}\right) \right)\Big|_{t
= 0}\nonumber\\
& = \dd{t} \left(f \circ \pi \left(L(p) e^{t \mathrm{Ad}_{h(p)}
A}h(p)\right)
\right)\Big|_{t= 0}\nonumber\\
& = \left( \mathrm{Ad}_{h(p)} A\right)^\sharp\big|_{L(p)} [f \circ \pi]
\nonumber \\
\label{eqn:adhpi}& = \pi_* \big((\mathrm{Ad}_{h(p)} A)^\sharp\big|_{L(p)}
\big)[f].
\end{align}
Now expand $A$ in the basis of $\mathfrak{k}$ as $A=X^AK_A$, where the
$X^A$ are constant supernumbers of parity $A$. Then, as the function $f$
is arbitrary, and by appropriate choice of the $X^A$, we find
\begin{equation}\label{eqn:splodge}
\pi_* \big(K_A{}^\sharp\big|_{L(p)h(p)} \big) = \pi_*
\big((\mathrm{Ad}_{h(p)}K_A)^\sharp)\big|_{L(p)} \big).
\end{equation}
Let us introduce the coadjoint representation, $g\mapsto
 \Lambda_p{}^q(g)$, of the group $G$ defined by
\begin{equation}\label{eqn:coadjoint}
\mathrm{Ad}_{g^{-1}}T_p \defined  \Lambda_p{}^q(g)T_q.
\end{equation}
We will now again see the importance of the assumption that the group $G$
is
 reductive. In this case we have
$[\mathfrak{h},\mathfrak{k}]\subseteq\mathfrak{k}$ and thus the coadjoint
representation furnishes us with a representation, $h\mapsto
\Lambda_A{}^B(h)$, of the subgroup $H$ on the space $\mathfrak{k}$
\begin{equation}
\mathrm{Ad}_{h^{-1}}K_A =  \Lambda_A{}^B(h)K_B.
\end{equation}
Thus, from \refe{eqn:splodge}, we finally have
\begin{equation}\label{eqn:gaugetransformframe}
E_A^{(L h)} = \Lambda_A{}^B(h^{-1}) E_B^{(L)}.
\end{equation}

It follows immediately from \refe{eqn:gaugetransformframe} that the
coframe transforms under a gauge transformation as
\begin{equation}\label{eqn:gaugetransformcoframe}
E^A_{(L h)} =  E^B_{(L)} \Lambda_B{}^A(h).
\end{equation}
This can also be derived directly from the definition \refe{eqn:coframe}
using \refe{eqn:Xchangesection} in a similar manner to that used in the
case of the connection.

We are primarily, though not solely, interested in the case when $G/H$ is
a superspace, see Section \ref{sec:intro}. Recall that one crucial
property of superspace is that its tangent space group must coincide with
the even Grassmann shell \cite{bluebook} of the tangent space group of the
body of the superspace under consideration. This will be the case when $H$
is the even Grassmann shell of $SO(p,q)$. Further to this we require for a
superspace that the representation $\Lambda_A{}^B(h)$ is completely
reducible, acting as $SO(p,q)$ rotations on the even coordinates and as
$\mathrm{Spin}(p,q)$ rotations on the odd coordinates. In this case we
will sometimes refer to the frame as being orthonormal although this is
only true for the even part of the frame.

\section{Torsion and curvature}\label{sec:curvature}
In the previous sections we have prescribed a supergeometry on the coset
space $G/H$ by determining a local frame and connection.
\refec{eqn:localconnectiondefinition}{eqn:coframe} may be combined into
one equation
\begin{equation}\label{eqn:theequation}
L^\ast\zeta=E^A_{(L)}K_A+\omega^I_{(L)}H_I,
\end{equation}
where we have expanded the connection in the generators of the algebra
$\mathfrak{h}$. This equation is often stated in the literature
\cite{hep-th/0503196,lightblue} in a matrix representation so that
$L^\ast\zeta = L(p)^{-1}\extd L(p)$, see Section \ref{sec:maurercartan}.
Given a supergeometry it is natural to next calculate the torsion and
curvature. We will see that the torsion and curvature can be
straightforwardly determined from the Maurer-Cartan structure equation,
\refe{eqn:maurercartanstructure}.

\subsection{Local torsion and curvature}
First we define the torsion and curvature on the base. In this section we
will not explicitly display the dependence of the frame and connection on
the local section $L$, all quantities derived from them do, however,
obviously retain this gauge dependence. From the local connection $\omega$
we define the curvature on the base as
\begin{equation}\label{eqn:curvaturedef}
R \defined \extd \omega + \omega \wedge \omega,
\end{equation}
which can be expanded in components as $R=R^IH_I$. This gives
\begin{equation}
R^I = \extd \omega^I - \frac{1}{2} \omega^J \wedge \omega^K f_{KJ}{}^I.
\end{equation}
The torsion is naturally defined in terms of the exterior covariant
derivative of the frame\footnote{Note that the definition we use has the
  opposite sign to much of the literature.}
\begin{equation}\label{eqn:torsiondef}
T^A\defined -\extcov E^A =  - \extd E^A - E^B \wedge \omega_B{}^A.
\end{equation}
Here $\omega_B{}^A$ is the connection $\omega$ written in terms of the
coadjoint representation. Using the definition of the coadjoint
representation, \refe{eqn:coadjoint}, we see that to first order
\begin{equation}\label{eqn:infinitesimalcoadjoint}
\Lambda_A{}^B(1+\epsilon^IH_I) = \delta_A{}^B-\epsilon^If_{IA}{}^B,
\end{equation}
hence
\begin{equation}
\omega_A{}^B= - \omega^If_{IA}{}^B.
\end{equation}

We may also consider the component form of the definitions of curvature
and torsion. We expand $\omega=E^A\omega_A$, and similarly
$\omega_B{}^C=E^A\omega_{AB}{}^C$, and also introduce the \emph{anholonomy
supercoefficients}, $\mathcal{C}_{AB}{}^C$, via
\begin{equation}
[E_A,E_B]=\mathcal{C}_{AB}{}^CE_C.
\end{equation}
It is then possible to show that
\begin{subequations}
\begin{gather}
\label{eqn:comptorsion} T_{AB}{}^C = \mathcal{C}_{AB}{}^C +
\omega_{AB}{}^C -
(-1)^{AB}\omega_{BA}{}^C \\
\label{eqn:compcurvature}R_{AB} = E_A [\omega_B] - (-1)^{AB} E_B
[\omega_A] - \mathcal{C}_{AB}{}^C\omega_C + [\omega_A,\omega_B].
\end{gather}
\end{subequations}
Using the conventions of Appendix \ref{sec:formconventions} we can also
show that
\begin{equation}\label{eqn:RXY}
R(X,Y)=(-1)^{XY}\big(X[\omega(Y)] - (-1)^{XY}Y[\omega(X)] - \omega([X,Y])
+ [\omega(X),\omega(Y)]\big)
\end{equation}
for supervectors $X$ and $Y$. These expressions will be useful later.

With these definitions in mind let us consider the pullback of the
Maurer-Cartan structure equation, \refe{eqn:maurercartanstructure}, under
the local section. We have
\begin{displaymath} \extd L^\ast \zeta +  L^\ast \zeta
\wedge  L^\ast \zeta = 0.
\end{displaymath}
Substituting in \refe{eqn:theequation} and separating out the
$\mathfrak{k}$ and $\mathfrak{h}$ parts we find
\begin{align*}
\extd E^A - \frac{1}{2}E^B \wedge E^C f_{CB}{}^A - E^B \wedge \omega^I
f_{IB}{}^A &= 0 \\
\extd \omega^I - \frac{1}{2}E^B \wedge E^C f_{CB}{}^I - \frac{1}{2}
\omega^J \wedge \omega^K f_{KJ}{}^I &= 0.
\end{align*}
Comparing these two equations with the definitions of the curvature and
torsion two-forms we see that
\begin{subequations}
\begin{align}
\label{eqn:torsioncomponent} T^A &= - \frac{1}{2}E^B \wedge E^C f_{CB}{}^A \\
\label{eqn:curvaturecomponent} R^I &= \frac{1}{2}E^B \wedge E^C
f_{CB}{}^I.
\end{align}
\end{subequations}
Comparing to the component expansion of a $p$-form as given in
\refe{eqn:formexpansion} we see that
\begin{equation}\label{eqn:TRcomponents}
T_{AB}{}^C=f_{AB}{}^C, \qquad R_{AB}{}^I=-f_{AB}{}^I.
\end{equation}

\subsection{Torsion and curvature from the bundle}
Whilst it is sufficient for most applications to consider the torsion and
curvature as defined in the previous section, it is nice to see how they
are obtained in terms of the geometry of the bundle. The first concept we
need to define is the exterior covariant derivative, $\extcov$, on the
bundle, not to be confused with that defined on the base used in the
previous section. It is defined by its action on an $n$-form $\phi$ on the
bundle as
\begin{equation}
\extcov \phi (X_1, \ldots , X_{n+1}) \defined \extd \phi (X^\mathrm{h}_1,
\ldots , X^\mathrm{h}_{n+1}).
\end{equation}
Here the supervector $X^\mathrm{h}$ is the horizontal part of the
supervector $X$.

Now recall that the vertical part of the Maurer-Cartan form gave us the
connection on the bundle $\zeta^IH_I=\Omega$. Denoting the horizontal part
by $\zeta^AK_A=\Theta$, we have
\begin{equation}\label{eqn:zetasplit2}
\zeta=\Theta+\Omega.
\end{equation}
The form $\Theta$ is essentially the \emph{solder form} on the bundle
\cite{goeckelerschuecker} which can be seen as follows. The solder form is
a form defined on a frame bundle; we will see in Section \ref{sec:equiv}
how the principal bundle may be thought of as the frame bundle. In
particular, a frame $E^{(L)}_A|_p$ corresponds to the point $L(p)$ in the
principal bundle. Given a vector $V|_{L(p)}$ tangent to the bundle at this
point the components of the solder form, $\Theta^A$, are defined to
satisfy
\begin{equation}
\pi_\ast \left(V|_{L(p)}\right) = \Theta^A(V|_{L(p)}) E^{(L)}_A|_p.
\end{equation}
By expanding $V$ in the basis of the $T_p^\sharp$ it is straightforward to
check that $\Theta^A=\zeta^A$ solves this equation. $\Theta=\zeta^AK_A$
can therefore be considered as the collection of the components of the
solder form into a single algebra valued form, thus in the following we
will simply refer to $\Theta$ as the solder form.

We may then define the quantities $\mathcal{T}$ and $\mathcal{R}$ which we
call the torsion and curvature on the bundle. These are defined in terms
of the exterior covariant derivatives of the solder form and connection
form, respectively, and are calculated to be
\begin{subequations}
\begin{align}
\label{eqn:bundletorsion}-\mathcal{T} &\defined \extcov \Theta = \extd
\Theta + \Theta \wedge \Omega + \Omega \wedge \Theta \\
\label{eqn:bundlecurvature}\mathcal{R} &\defined \extcov \Omega = \extd
\Omega + \Omega \wedge \Omega.
\end{align}
\end{subequations}
Note that $\mathcal{T}$ is a $\mathfrak{k}$-valued two-form, whereas
$\mathcal{R}$ is $\mathfrak{h}$-valued. A straightforward calculation
shows that when pulled back under a local section these quantities give
the definitions of the local torsion and curvature two-forms in
\refec{eqn:curvaturedef}{eqn:torsiondef}
\begin{equation}\label{eqn:pullbackct}
L^\ast \mathcal{T}=T^AK_A, \qquad L^\ast \mathcal{R}=R.
\end{equation}

Let us now consider how the torsion and curvature on the bundle may be
expressed in terms of the structure constants of the algebra. Using
\refe{eqn:zetasplit2} and \refec{eqn:bundletorsion}{eqn:bundlecurvature}
we see that
\begin{align}
\nonumber \extcov \zeta = -\mathcal{T}+\mathcal{R} &= \extd \Theta +
\Theta \wedge \Omega + \Omega \wedge \Theta + \extd \Omega + \Omega \wedge
\Omega \\
\nonumber &= \extd \zeta + \zeta \wedge \zeta - \Theta \wedge \Theta \\
&= - \Theta \wedge \Theta,
\end{align}
where in the last line we have used the Maurer-Cartan structure equation
\refe{eqn:maurercartanstructure}. As $\Theta=\zeta^AK_A$ we find
\begin{align*}
\nonumber -\mathcal{T}+\mathcal{R} &= \frac{1}{2}\zeta^A \wedge \zeta^B
 [K_B,K_A] \\
&= \frac{1}{2}\zeta^A \wedge \zeta^B f_{BA}{}^C K_C +
 \frac{1}{2}\zeta^A \wedge \zeta^B f_{BA}{}^I H_I.
\end{align*}
If we then split this up into its $\mathfrak{k}$ and $\mathfrak{h}$ parts
we find
\begin{subequations}
\begin{align}
\mathcal{T} &= -\frac{1}{2}\zeta^A \wedge \zeta^B f_{BA}{}^C K_C \\
\mathcal{R} &= \frac{1}{2}\zeta^A \wedge \zeta^B f_{BA}{}^I H_I.
\end{align}
\end{subequations}
If we pullback these equations to the base as in \refe{eqn:pullbackct} we
obtain the expressions for the local torsion and curvature we derived
earlier in \refec{eqn:torsioncomponent}{eqn:curvaturecomponent}.

\section{Associated bundles and supertensor bundles}
\label{sec:bundle}
\subsection{Associated bundles}
\label{sec:associated} Given a principal bundle $P(G/H,H) = G$ we
construct an associated fibre bundle as follows. Consider a manifold $F$
on which the structure group $H$ acts on the left. Then we define an
equivalence relation on $G \times F$ by
\begin{equation}
\label{eqn:equivrel} (g,f) \sim (gh,h^{-1}f),
\end{equation}
where $(g,f) \in G \times F$. In the following we shall denote the
equivalence class of the point $(g,f)$ as $[(g,f)]$. From this equivalence
relation we define the associated fibre bundle as the coset space $(G
\times F)/H$, see \cite{nakahara}. In the following we shall consider
associated bundles only where the fibre $F$ is a supervector
space\footnote{Note that local sections of associated bundles where the
fibre $F$ is a supervector space can be linearly combined to give new
local sections in a way that will become apparent in Section
\ref{sec:supertensorbundles}.}. In these cases $H$ acts on the fibre via
some representation $\rho$ and the equivalence relation,
\refe{eqn:equivrel}, reads
\begin{equation}
\label{eqn:equivrelvector} (g,\xi) \sim (gh,\rho(h)^{-1}\xi), \quad \xi
\in F.
\end{equation}

Defining the local bundle map
\begin{equation}
\label{eqn:defRhL} R_{h_1}^{(L)}  : L(p)h  \hspace{5pt} \mapsto
\hspace{5pt}  L(p)h_1 h
\end{equation}
we can define the action of $H$ on a local section of a general associated
bundle $s(p) = [(L(p),\xi(p))]$ as
\begin{equation}
\label{eqn:tildeRhL} \tilde{R}_h^{(L)}  : [(L(p), \xi(p))]  \hspace{5pt}
\mapsto \hspace{5pt}  [(R_h^{(L)}\big(L(p)\big), \xi(p))]
\end{equation}
which, using the equivalence relation, can be rewritten as
\begin{equation}
\label{eqn:hactionsection} \tilde{R}_{h}^{(L)} s(p) = [(L(p), \rho(h)
\xi(p))].
\end{equation}
Note that $R_{h}^{(L)}$ is a group homomorphism, i.e.\ we have
\begin{equation}
R_{h_1}^{(L)}\circ R_{h_2}^{(L)} = R_{h_1h_2}^{(L)}.
\end{equation}

Based on the definition of $\tilde{R}_h^{(L)}$ we can define the action of
$\mathfrak{h}$ on a local section $s(p) = [(L(p),\xi(p))]$ of a general
associated bundle. For $h_\epsilon = 1 + \epsilon H +
\mathcal{O}(\epsilon^2)$ we define
\begin{equation}
\label{eqn:deftildeRH} \tilde{R}_H^{(L)} s(p) \defined \dd{t}\Big(
\tilde{R}_{h_t}^{(L)} s(p)\Big) \Big|_{t = 0} .
\end{equation}
We hence have
\begin{equation}
\tilde{R}_H^{(L)} s(p) = \Big\lbrack \Big(L(p),
\dd{t}(\rho(h_t)\xi(p))\Big|_{t = 0}\Big) \Big\rbrack.
\end{equation}

\subsection{Equivalence of the principal bundle with the frame bundle}
\label{sec:equiv} In this section we shall show that the
orthonormal\footnote{Note that here we use the term orthonormal in the
sense discussed in the last paragraph of Section \ref{sec:frame}.} frame
bundle $F(G/H)$ is equivalent to the principal bundle. Note that the fibre
of the frame bundle above the point $p$ is given by $\{E_A^{(L_i
h)}\big|_p: h\in H \} \cong H$. We define the local trivialization of the
frame bundle as
\begin{align}
\psi_i&:U_i\times H \to \pi_F^{-1}(U_i) \nonumber \\
&:(p,h)\mapsto \{\Lambda_A{}^B(h^{-1}) E_B^{(L_i)}\big|_p \}= \{E_A^{(L_i
h)}\big|_p\},
\end{align}
where we have defined the frame bundle projection map $\pi_F$ as
\begin{align}
\pi_F &: F(G/H) \to G/H\nonumber \\
&: \pi_* \big(K_A{}^\sharp\big|_g \big) \mapsto gH.
\end{align}
The transition functions $\tilde{t}_{ij}$ of the frame bundle are then
given by
\begin{equation}
\tilde{t}_{ij}(p)=\psi_{i,p}^{-1}\psi_{j,p}:H\to H,
\end{equation}
where $\psi_i(p,h)=\psi_{i,p}(h)$. We thus already see that the structure
group of the orthonormal frame bundle is equal to the structure group of
the principal bundle, see Section \ref{sec:principal}. In order to fully
establish the equivalence between the principal bundle and the frame
bundle we shall now show that the transition functions of the principal
bundle are equal to those of the frame bundle. In order to do this
consider the action of $\tilde{t}_{ij}(p)$ on $h$
\begin{equation}
\tilde{t}_{ij}(p)h =
\psi_{i,p}^{-1}\left(\Lambda_A{}^B(h^{-1})E_B^{(L_j)}\big|_p\right).
\end{equation}
Now we have using the definition of the transition functions of the
principal bundle $P(G/H,H)$, see \refe{eqn:localtriv},
\begin{equation}
L_i(p) = \phi_i(p,e)= \phi_j(p, t_{ji}(p)e) = L_j(p)t_{ji}(p)
\end{equation}
and hence we find
\begin{align}
E_B^{(L_j)}\big|_p & = \pi_* \big(K_B{}^\sharp\big|_{L_j(p)} \big)\nonumber \\
& = \pi_* \big(K_B{}^\sharp\big|_{L_i(p)t_{ij}} \big) \nonumber\\
& = \Lambda_B{}^C(t_{ij}(p)^{-1})E_C^{(L_i)}\big|_p.
\end{align}
We thus have
\begin{align}
\tilde{t}_{ij}(p)h & =
\psi_{i,p}^{-1}\left(\Lambda_A{}^B(h^{-1})\Lambda_B{}^C(t_{ij}(p)^{-1})E_C^{(L_i)}\big|_p\right)
\nonumber \\
& =
\psi_{i,p}^{-1}\left(\Lambda_A{}^B\left((t_{ij}(p)h)^{-1}\right)E_B^{(L_i)}\big|_p\right)
\nonumber \\
& = t_{ij}(p)h.
\end{align}
This proves that the transition functions of the frame bundle equal those
of the principal bundle. We thus see that the principal bundle and the
frame bundle are equivalent bundles. In the following we shall, for
convenience, use the principal bundle rather than the frame bundle in
order to formulate associated bundles.

\subsection{Supertensor bundles}
\label{sec:supertensorbundles} In this section we shall briefly discuss
the equivalence relations \refe{eqn:equivrel} in the case of general
tensor bundles. Although it is more natural to consider tensor bundles
associated to the frame bundle we will in this section use the equivalent
principal bundle for the ease of notation.

Let us consider a pure, i.e.\ even or odd, section $s(p)$ of a general
tensor bundle with both contravariant and covariant indices. We write
\begin{equation}
\label{eqn:supertensor} s(p) = [(L(p), \{s^{A_1 \cdots
A_i}{}_{\tilde{A}_{i + 1} \cdots \tilde{A}_{i + j}}{}^{A_{i + j + 1}
\cdots  }\cdots\})].
\end{equation}
Here we define the equivalence by
\begin{multline}
\label{eqn:equivreltensor1} (L(p),\{s^{A_1 \cdots A_i}{}_{\tilde{A}_{i +
1} \cdots \tilde{A}_{i + j}}{}^{A_{i + j + 1} \cdots }\cdots \})
\\ \sim (L(p)h,\rho(h^{-1})\{s^{A_1 \cdots A_i}{}_{\tilde{A}_{i +
1} \cdots \tilde{A}_{i + j}}{}^{A_{i + j + 1} \cdots  }\cdots \})
\end{multline}
where we set
\begin{multline}
\label{eqn:equivreltensor2} \rho(h)\{s^{A_1 \cdots A_i}{}_{\tilde{A}_{i +
1} \cdots \tilde{A}_{i + j}}{}^{A_{i + j + 1} \cdots }\cdots\} \\ =
\{(-1)^{\Delta_n(A+B,B + \tilde{B}) + \Delta_n(A + \tilde{B},\tilde{A} +
\tilde{B})} \Lambda_{\tilde{A}_{i + 1}}{}^{\!\!\!\tilde{B}_{i + 1}}(h)
\cdots \Lambda_{\tilde{A}_{i + j}}{}^{\!\!\!\tilde{B}_{i + j}}(h)\cdots  \\
s^{B_1 \cdots B_i}{}_{\tilde{B}_{i + 1} \cdots \tilde{B}_{i + j}}{}^{B_{i
+ j + 1} \cdots }\cdots  \\ \Lambda_{B_1}{}^{\!\!\!A_1}(h^{-1}) \cdots
\Lambda_{B_i}{}^{\!\!\!A_i}(h^{-1})  \Lambda_{B_{i + j + 1}}{}^{\!\!\!A_{i
+ j + 1}}(h^{-1}) \cdots\}.
\end{multline}
Here $\Delta_n$ is the parity function as defined in
\refe{eqn:parityfunction}, $n$ denotes the total number of indices and we
have set $\Delta_n(A,B + \tilde{B}) \equiv \Delta_n(A,B) +
\Delta_n(A,\tilde{B})$.

We define the local section of a general supertensor bundle to satisfy the
following linearity property under left multiplication by an arbitrary
pure supernumber $\lambda$
\begin{equation}
\label{eqn:linearity} \lambda s(p)  \defined [(L(p),\{ (-1)^{\lambda (\sum
\tilde{A})}\lambda s^{A_1 \cdots A_i}{}_{\tilde{A}_{i + 1} \cdots
\tilde{A}_{i + j}}{}^{A_{i + j + 1} \cdots }\cdots\})],
\end{equation}
where $\sum \tilde{A} \equiv ( \tilde{A}_{i + 1} + \ldots + \tilde{A}_{i +
j} + \ldots)$ is the sum over the parities of the lower indices. From
\refe{eqn:linearity} and from the fact that $\lambda s(p) = (-1)^{\lambda
s}s(p)\lambda$ we can infer the linearity property under right
multiplication
\begin{equation}
s(p) \lambda = [(L(p),\{ (-1)^{\lambda (\sum A)}s^{A_1 \cdots
A_i}{}_{\tilde{A}_{i + 1} \cdots \tilde{A}_{i + j}}{}^{A_{i + j + 1}
\cdots  }\cdots \lambda \})].
\end{equation}
From these last two equations we see that contravariant tensor bundles are
left-linear, i.e.\ they satisfy
\begin{equation}
\lambda s(p) = [(L(p),\{\lambda s^{A_1 \cdots A_n} \})],
\end{equation}
whereas covariant tensor bundles are right-linear, i.e.
\begin{equation}
s(p) \lambda = [(L(p),\{s_{A_1 \cdots A_n}\lambda \})].
\end{equation}
Defining the basis of a general supertensor bundle by\footnote{One should
note that the collection of $\delta$'s that occurs in the definition of
the basis is to be understood as an ordered collection, i.e.\ the
$\delta$'s ought not be swapped.}
\begin{multline}
\label{eqn:basisdef} E_{A_1}^{(L)} \otimes \cdots \otimes E_{A_i}^{(L)}
\otimes E^{\tilde{A}_{i + 1}}_{(L)} \otimes \cdots \otimes E^{\tilde{A}_{i
+ j}}_{(L)} \otimes E_{A_{i+ j + 1}}^{(L)} \otimes \cdots
\defined [(L(p),\\ \{(-1)^{\Delta_n(A,A) + \Delta_n(\tilde{A},\tilde{A}) }
\delta_{A_1}{}^{\!\!\!B_1}\cdots\delta_{A_i}{}^{\!\!\!B_i}
\delta_{\tilde{B}_{i + 1}}{}^{\!\!\!\tilde{A}_{i +
1}}\cdots\delta_{\tilde{B}_{i + j}}{}^{\!\!\!\tilde{A}_{i + j}}
\delta_{A_{i + j + 1}}{}^{\!\!\!B_{i + j + 1}}\cdots\})]
\end{multline}
we can rewrite the local section $s(p)$ of \refe{eqn:supertensor} in terms
of this basis as
\begin{align}
s(p) & = [(L(p), \{s^{A_1 \cdots A_i}{}_{\tilde{A}_{i + 1} \cdots
\tilde{A}_{i + j}}{}^{A_{i + j + 1} \cdots  }\cdots\})] \nonumber\\
& = (-1)^{\Delta_n(A,A) + \Delta_n(\tilde{A},\tilde{A})}(-1)^{(s + \sum A
+ \sum \tilde{A})(\sum \tilde{A})} s^{A_1 \cdots A_i}{}_{\tilde{A}_{i + 1}
\cdots \tilde{A}_{i + j}}{}^{A_{i + j + 1} \cdots }\cdots \nonumber \\
& \hspace{15pt} E_{A_1}^{(L)} \otimes \cdots \otimes E_{A_i}^{(L)} \otimes
E^{\tilde{A}_{i + 1}}_{(L)} \otimes \cdots \otimes E^{\tilde{A}_{i +
j}}_{(L)} \otimes E_{A_{i+ j + 1}}^{(L)} \otimes \cdots \nonumber \\ & =
(-1)^{\Delta_n(A + \tilde{A},A + \tilde{A})}(-1)^{(s + \sum
\tilde{A})(\sum \tilde{A})} s^{A_1 \cdots A_i}{}_{\tilde{A}_{i + 1} \cdots
\tilde{A}_{i + j}}{}^{A_{i + j + 1} \cdots }\cdots \nonumber\\ &
\hspace{15pt} E_{A_1}^{(L)} \otimes \cdots \otimes E_{A_i}^{(L)} \otimes
E^{\tilde{A}_{i + 1}}_{(L)} \otimes \cdots \otimes E^{\tilde{A}_{i +
j}}_{(L)} \otimes E_{A_{i+ j + 1}}^{(L)} \otimes \cdots, \nonumber
\end{align}
where we have used the linearity property \refe{eqn:linearity}. Note that
in the case of contravariant tensor bundles this formula simplifies to
\begin{equation}
s(p) = (-1)^{\Delta_n(A,A)} s^{A_1 \cdots A_n} E_{A_1}^{(L)} \otimes
\cdots \otimes E_{A_n}^{(L)}
\end{equation}
and in the case of covariant tensor bundles we have
\begin{equation}
\label{eqn:covariantexp} s(p) = (-1)^{\Delta_n(A,A)}E^{A_1}_{(L)} \otimes
\cdots \otimes E^{A_n}_{(L)}s_{A_1 \cdots A_n}.
\end{equation}

Using the definition of the action of $H$ on a local section $s(p)$ as
given in \refe{eqn:hactionsection} we find after a slightly lengthy
calculation for the transformation of the basis under $H$
\begin{multline}
\tilde{R}_{h}^{(L)} \left(E_{A_1}^{(L)} \otimes \cdots \otimes
E_{A_i}^{(L)} \otimes E^{\tilde{A}_{i + 1}}_{(L)} \otimes \cdots \otimes
E^{\tilde{A}_{i +
j}}_{(L)} \otimes E_{A_{i+ j + 1}}^{(L)} \otimes \cdots \right) \\
= E_{A_1}^{(Lh)} \otimes \cdots \otimes E_{A_i}^{(Lh)} \otimes
E^{\tilde{A}_{i + 1}}_{(Lh)} \otimes \cdots \otimes E^{\tilde{A}_{i +
j}}_{(Lh)} \otimes E_{A_{i+ j + 1}}^{(Lh)} \otimes \cdots
\end{multline}
as expected.

From the definition of the basis \refe{eqn:basisdef} and the linearity
property \refe{eqn:linearity} one can easily deduce the following
properties of the tensor product
\begin{subequations}
\begin{align}
\lambda (s_1 \otimes s_2) & = (\lambda s_1) \otimes s_2  \equiv \lambda
s_1 \otimes s_2\\
(s_1 \otimes s_2)\lambda  & = s_1 \otimes (s_2 \lambda)  \equiv s_1
\otimes s_2 \lambda\\
(s_1 \lambda )\otimes s_2 & = s_1 \otimes (\lambda s_2)  \equiv s_1
\lambda \otimes s_2 \\
(s_1 \otimes s_2) \otimes s_3 & = s_1 \otimes (s_2 \otimes s_3) \equiv s_1
\otimes s_2 \otimes s_3,
\end{align}
\end{subequations}
where $s_1$, $s_2$ and $s_3$ are local sections of general tensor bundles
and $\lambda$ is a supernumber.

\section{Bundle maps}
Let us define the following (local) bundle maps
\begin{subequations}
\begin{alignat}{2}
\label{eqn:defLg}
L_g & : L(p)h & \hspace{5pt} \mapsto \hspace{5pt}  & g L(p) h\\
L_g^{(L)} & : L(p)h & \hspace{5pt} \mapsto \hspace{5pt} & g L(p) \tilde{h}_L^{(L)}(p,g)^{-1}h \\
\label{eqn:defRgL} R_g^{(L)} & : L(p)h & \hspace{5pt} \mapsto \hspace{5pt}
& L(p)gh.
\end{alignat}
\end{subequations}
While $L_g$, see Section \ref{sec:invvecfields}, is a global bundle map
from $G \rightarrow G$ -- here written locally in the patch $\pi^{-1}(U)$
-- the map $L_g^{(L)}$ depends on the local section $L(p)$ and, for it to
be well-defined, both its domain and range must be restricted to
$\pi^{-1}(U)$. In the following we shall however, for convenience, also
restrict both the domain and range of the map $L_g$ to $\pi^{-1}(U)$. Then
the \emph{left} $H$-compensator $\tilde{h}_L^{(L)}(p,g)$ is defined by the
equation, see \cite{lightblue,hep-th/0007099,hep-th/9912277},
\begin{subequations}
\begin{equation}
\label{eqn:leftHcomp} g L(p) = L(q) \tilde{h}_L^{(L)}(p,g) \quad
\mathrm{with} \, p,q \in U,
\end{equation}
i.e., we have
\begin{equation}
\label{eqn:LgL} L_g^{(L)}\left(L(p)h\right)= L(q)h
\end{equation}
and
\begin{equation}
\label{eqn:Lg} L_g \left(L(p)h\right)= L(q)\tilde{h}_L^{(L)}(p,g) h
\end{equation}
\end{subequations}
and hence we see that the map $L_g^{(L)}$ preserves the local section
$L(p)$, while this is not true for the map $L_g$.

For the map $R_g^{(L)}$ -- which also depends on the local section $L(p)$
-- only the domain need be restricted to $\pi^{-1}(U)$; in the following,
however, we shall for convenience also require its range to be restricted
to $\pi^{-1}(U)$. We can then, in accordance with \refe{eqn:leftHcomp},
also define the \emph{right} $H$-compensator $\tilde{h}_R^{(L)}(p,g)$ by
\begin{subequations}
\begin{equation}
\label{eqn:rightHcomp} L(p) g = L(q) \tilde{h}_R^{(L)}(p,g), \quad
\mathrm{with} \, p,q \in U,
\end{equation}
i.e.,
\begin{equation}
\label{eqn:RgL} R_g^{(L)}\left(L(p)h\right)= L(q)\tilde{h}_R^{(L)}(p,g)h.
\end{equation}
\end{subequations}
Note that the map $R_h^{(L)}$ introduced in Section \ref{sec:associated},
see \refe{eqn:defRhL}, is just a specific case of the map $R_g^{(L)}$
introduced here.

The maps $L_g^{(L)}$ and $R_g^{(L)}$ induce maps on the base that we shall
denote by $l_g$ and $r^{(L)}_g$, respectively. We have
\begin{subequations}
\begin{align}
\label{eqn:piL}
\pi \circ L_g^{(L)} & = \pi \circ L_g = l_g \circ \pi\\
\label{eqn:piR} \pi \circ R_g^{(L)} & = r_g^{(L)} \circ \pi,
\end{align}
\end{subequations}
where one should note that the map $r_g^{(L)}$ on the base depends on the
local section $L(p)$. We can then rewrite \refec{eqn:LgL}{eqn:Lg} and
\refe{eqn:RgL}, respectively, as
\begin{subequations}
\begin{align}
L_g^{(L)}\left(L(p)h\right) & = L(l_g(p))h \\
L_g \left(L(p)h\right) & = L(l_g(p))\tilde{h}_L^{(L)}(p,g)h \\
R_g^{(L)}\left(L(p)h\right) & = L(r_g^{(L)}(p))\tilde{h}_R^{(L)}(p,g)h.
\end{align}
\end{subequations}
Using \refec{eqn:piL}{eqn:piR} we can derive relations for the vectors
$\pi_* (A^\sharp|_{L(p)})$ and $\pi_* (A^\flat|_{L(p)})$ similar to those
in \refec{eqn:sharp}{eqn:flat}. We have
\begin{align}
\pi_* \left(A^\sharp\big|_{L(p)} \right)[f] & = A^\sharp \big|_{L(p)} [f
\circ \pi] \nonumber \\
& = \dd{t}\left(f \circ \pi\left(L(p) e^{tA}\right) \right)\Big|_{t=0}\nonumber \\
& = \dd{t} \left( f \circ \pi \circ R_{e^{tA}}^{(L)}L(p)  \right)
\Big|_{t=0} \nonumber \\
& = \dd{t} \left( f \circ r_{e^{tA}}^{(L)} \circ \pi \circ L(p)  \right)
\Big|_{t=0} \nonumber \\ \label{eqn:expansionf(r)}& = \dd{t}
f(r_{e^{tA}}^{(L)}(p))\Big|_{t=0}
\end{align}
and similarly we find
\begin{equation}\label{eqn:pistarAflatdefinition}
\pi_* \left(A^\flat\big|_{L(p)} \right)[f] = \dd{t}
f(l_{e^{tA}}(p))\Big|_{t=0}.
\end{equation}
Note that the right hand side of this last equation is independent of the
local section $L(p)$, and hence we have that
\begin{equation}\label{eqn:notmultiplydefined}
\pi_* \left(A^\flat\big|_{L(p)} \right) = \pi_*
\left(A^\flat\big|_{L(p)h(p)} \right) .
\end{equation}
Thus we may unambiguously write $\pi_\ast A^\flat$ as a well defined
vector field on $G/H$ without the need to specify which point in the fibre
the supervector $A^\flat$ was based. Note that this not true for
$A^\sharp$.

The maps $L_g$, $L_g^{(L)}$ and $R_g^{(L)}$ defined in
\refes{eqn:defLg}{eqn:defRgL}, respectively, can be extended to maps on
associated bundles as follows. Consider a local section $s(p) = [(L(p),
\xi(p))]$ of some general associated bundle. We then define the maps
$\tilde{L}_g$, $\tilde{L}_g^{(L)}$ and $\tilde{R}_g^{(L)}$ by
\begin{subequations}
\begin{alignat}{2}
\label{eqn:tildeLg}
\tilde{L}_g & : [(L(p), \xi(p))] & \hspace{5pt} \mapsto \hspace{5pt} & [(L_g\big(L(p)\big), \xi(p))]\\
\tilde{L}_g^{(L)} & : [(L(p), \xi(p))] & \hspace{5pt} \mapsto \hspace{5pt} & [(L_g^{(L)}\big(L(p)\big), \xi(p))]\\
\label{eqn:tildeRgL} \tilde{R}_g^{(L)} & : [(L(p), \xi(p))] & \hspace{5pt}
\mapsto \hspace{5pt} & [(R_g^{(L)}\big(L(p)\big), \xi(p))].
\end{alignat}
\end{subequations}

\subsection{Properties of the $H$-compensators}\label{sec:propertiesHcompensators}
In this section we will discuss properties of the left and right
$H$-compensators.
\subsubsection{Properties of the left $H$-compensator}
Recall that the left $H$-compensator is defined by the equation
\begin{equation}\label{eqn:lgproperty}
L(l_g(p)) = g L(p) \tilde{h}_L^{(L)}(p,g)^{-1}.
\end{equation}
Now, setting $l_g(p) = q$ we can rearrange this formula to give
\begin{displaymath}
L(p) = g^{-1} L(q)\tilde{h}_L^{(L)}(l_{g^{-1}}(q),g),
\end{displaymath}
but we also have
\begin{displaymath}
L(p)  = L(l_{g^{-1}}(q)) = g^{-1} L(q)\tilde{h}_L^{(L)}(q,g^{-1})^{-1}
\end{displaymath}
from which we can deduce the relation
\begin{equation}\label{eqn:tildehlrelinverse}
\tilde{h}_L^{(L)}(l_{g^{-1}}(q),g) = \tilde{h}_L^{(L)}(q,g^{-1})^{-1}.
\end{equation}
This relation will be important when considering derivations on local
sections of associated bundles.

Next we shall derive the composition rule for the left $H$-compensators.
We have
\begin{subequations}
\begin{align}
\label{eqn:lg1lg2} L^{(L)}_{g_1} \circ L^{(L)}_{g_2}\left(L(p)h \right) &
= L^{(L)}_{g_1} (L(l_{g_2}(p))
h) = L(l_{g_1} \circ l_{g_2}(p)) h \\
& = g_1 L(l_{g_2}(p)) \tilde{h}^{(L)}_L(l_{g_2}(p), g_1)^{-1}h
\nonumber\\\label{eqn:complefth1} & = g_1 g_2
L(p)\tilde{h}^{(L)}_L(p,g_2)^{-1}\tilde{h}^{(L)}_L(l_{g_2}(p), g_1)^{-1}
h.
\end{align}
\end{subequations}
On the other hand we have
\begin{subequations}
\begin{align}
\label{eqn:lg1g2} L^{(L)}_{g_1g_2}\left(L(p)h \right)
& = L(l_{g_1 g_2}(p))h \\
\label{eqn:complefth2} & =  g_1 g_2 L(p) \tilde{h}^{(L)}_L(p, g_1
g_2)^{-1}h .
\end{align}
\end{subequations}
Now, from \refec{eqn:complefth1}{eqn:complefth2}, we see that
\begin{equation}
\pi \circ L^{(L)}_{g_1g_2}\left(L(p)h \right) = \pi \circ L^{(L)}_{g_1}
\circ L^{(L)}_{g_2}\left(L(p)h \right)
\end{equation}
and hence we can deduce from \refec{eqn:lg1lg2}{eqn:lg1g2} and
\refec{eqn:complefth1}{eqn:complefth2}, respectively,
\begin{subequations}
\begin{align}
\label{eqn:complg's}
l_{g_1 g_2} & =  l_{g_1} \circ l_{g_2} \\
\label{eqn:comprulelefth}
 \tilde{h}^{(L)}_L(p, g_1 g_2)^{-1} & =
\tilde{h}^{(L)}_L(p,g_2)^{-1}\tilde{h}^{(L)}_L(l_{g_2}(p), g_1)^{-1} .
\end{align}
\end{subequations}
Now we shall also consider the transformation of $\tilde{h}^{(L)}_L$ under
a change of local section $L \rightarrow L' = Lh$, i.e.\ under a gauge
transformation. We have
\begin{subequations}
\begin{align}
\label{eqn:leftHcompgauge1}
g L'(p) \tilde{h}_L^{(L')}(p,g)^{-1} & = L'(l_g(p))\\
& = L(l_g(p)) h(l_g(p)) \nonumber \\
\label{eqn:leftHcompgauge2} & = g L(p) h(p)
h(p)^{-1}\tilde{h}_L^{(L)}(p,g)^{-1} h(l_g(p)).
\end{align}
\end{subequations}
From \refec{eqn:leftHcompgauge1}{eqn:leftHcompgauge2} we thus find
\begin{equation}
\label{eqn:tildehgauge} \tilde{h}_L^{(Lh)}(p,g)^{-1} =
h(p)^{-1}\tilde{h}_L^{(L)}(p,g)^{-1} h(l_g(p)).
\end{equation}
In the following we shall consider infinitesimal versions of
\refec{eqn:complg's}{eqn:comprulelefth} and \refe{eqn:tildehgauge}.
Defining
\begin{equation}
\label{eqn:defW} W^{(L)}_L(p,A) \defined - \frac{\mathrm{d}}{\mathrm{d}t}
\tilde{h}^{(L)}_L(p, e^{tA}) \Big|_{t = 0},
\end{equation}
we can write the expansion of $\tilde{h}^{(L)}_L(p,g)$ for $g =
e^{\epsilon A}$ with $\epsilon \ll 1$ and $A \in \mathfrak{g}$ to first
order as
\begin{equation}
\label{eqn:expansionhl} \tilde{h}_L^{(L)}(p,e^{\epsilon A}) =  1 -
\epsilon W^{(L)}_L (p,A) + \mathcal{O}(\epsilon^2).
\end{equation}
Note that using \refe{eqn:comprulelefth} we easily see that
$W^{(L)}_L(p,A)$ is left linear in $A$. In the following we shall for the
ease of notation suppress the $p$ dependence in $W^{(L)}_L$.

Now using the composition rule \refe{eqn:complg's} and the fact that
$(\mathrm{Ad}_{g_1} g_2) g_1 = g_1 g_2$ we find for the infinitesimal
version of \refe{eqn:complg's} with $g_1 = e^{\epsilon_1 A}$ and $g_2 =
e^{\epsilon_2 B}$ for $A, B \in \mathfrak{g}$
\begin{equation}\label{eqn:algebrapiflats}
[\pi_* A^\flat,\pi_* B^\flat] = - \pi_* [A,B]^\flat.
\end{equation}
This result can also be seen as a consequence of the well definedness of
the supervector field $\pi_\ast A^\flat$, \refe{eqn:notmultiplydefined},
which allows us to take the action of the pushforward under the
projection, $\pi_\ast$, inside the Lie bracket of \refe{eqn:flatalgebra}.

Now consider \refe{eqn:comprulelefth}. We have for the expansion of
$\tilde{h}^{(L)}_L(l_{g_2}(p), g_1)^{-1}$
\begin{displaymath}
\tilde{h}^{(L)}_L(l_{e^{\epsilon_2 B}}(p),e^{\epsilon_1 A} )^{-1}  = 1 +
\epsilon_1 W^{(L)}_L(A) + \epsilon_1\epsilon_2 \pi_* B^\flat
[W^{(L)}_L(A)]+\ldots .
\end{displaymath}
The infinitesimal version of \refe{eqn:comprulelefth} is thus given by
\begin{equation}
\label{eqn:intcondW1} W^{(L)}_L([A,B]) = \pi_* B^\flat [W^{(L)}_L(A)] -
\pi_* A^\flat [W^{(L)}_L(B)] + [W^{(L)}_L(B),W^{(L)}_L(A)].
\end{equation}
In the literature this last equation is normally referred to as the
integrability condition for the left $H$-compensator \cite{lightblue}.

Expanding \refe{eqn:tildehgauge} we find for the transformation of
$W^{(L)}_L$ under a gauge transformation $L \rightarrow Lh$
\begin{equation}
\label{eqn:gaugetrafoW} W^{(Lh)}_L(A) = h(p)^{-1} W^{(L)}_L(A) h(p) +
h(p)^{-1} \pi_*A^\flat h(p).
\end{equation}
Therefore we see that $W^{(L)}_L$ transforms like a connection under a
change of local section.

\subsubsection{Properties of the right $H$-compensator}
In a similar fashion as for the left $H$-compensator we can derive a
composition rule for the right $H$-compensator, \refe{eqn:rightHcomp},
$\tilde{h}^{(L)}_R(p,g)$. We find
\begin{subequations}
\begin{align}
\label{eqn:comprg's} r_{g_1}^{(L)} \circ r_{g_2}^{(L)}(p) & = r^{(L)}_{g_2
\mathrm{Ad}_{\tilde{h}_R^{(L)}(p,g_2)^{-1}}g_1}(p)\\
\label{eqn:comphr's} \tilde{h}_R^{(L)}(r_{g_2}^{(L)}(p),g_1)
\tilde{h}_R^{(L)}(p,g_2) & = \tilde{h}_R^{(L)}(p,g_2
\mathrm{Ad}_{\tilde{h}_R^{(L)}(p,g_2)^{-1}}g_1).
\end{align}
\end{subequations}
Again proceeding in a similar fashion as in the case of the left
$H$-compensator we can derive an expression for the transformation
properties of the right $H$-compensator under a change of local section $L
\rightarrow Lh$. Note, however, that in this case also the map $r_g^{(L)}$
will transform under a change of local section. We find
\begin{subequations}
\begin{align}
\label{eqn:gaugetransformrh}
r^{(Lh)}_{\mathrm{Ad}_{h(p)^{-1}}g}(p) & = r_g^{(L)}(p)\\
\label{eqn:gaugetransformrightHcomp}
\tilde{h}^{(Lh)}_R(p,\mathrm{Ad}_{h(p)^{-1}}g)^{-1} & =
h(p)^{-1}\tilde{h}_R^{(L)}(p,g)^{-1}h(r_g^{(L)}(p)).
\end{align}
\end{subequations}
In the following we shall consider the infinitesimal versions of
\refes{eqn:comprg's}{eqn:gaugetransformrightHcomp}. Defining
\begin{equation}
\label{eqn:defWR} W^{(L)}_R(p,A) \defined - \frac{\mathrm{d}}{\mathrm{d}t}
\tilde{h}^{(L)}_R(p, e^{tA}) \Big|_{t = 0},
\end{equation}
we can write the expansion of $\tilde{h}^{(L)}_R(p,g)$ for $g =
e^{\epsilon A}$ with $\epsilon \ll 1$ and $A \in \mathfrak{g}$ to first
order as
\begin{equation}
\label{eqn:expansionhr} \tilde{h}_R^{(L)}(p,e^{\epsilon A}) =  1 -
\epsilon W^{(L)}_R (p,A) + \mathcal{O}(\epsilon^2).
\end{equation}
Note that using \refe{eqn:comphr's} one can show that $W^{(L)}_R(p,A)$ is
left linear in $A$. In the following we shall for the ease of notation
suppress the $p$ dependence in $W^{(L)}_R$.

We shall now show that $W^{(L)}_R(A)$ for $A \in \mathfrak{k}$ is related
to the spin connection. In order to do this consider first
\begin{displaymath}
R^{(L)}_{e^{tK}} (L(p)) = L(p) e^{tK} =
L(r^{(L)}_{e^{tK}}(p))\tilde{h}_R^{(L)}(p,e^{tK}),
\end{displaymath}
for $K \in \mathfrak{k}$. Now, setting $\tilde{\gamma}(t) \equiv L(p)
e^{tK}$ and $\gamma(t) \equiv r^{(L)}_{e^{tK}}(p)$, we have $\pi \circ
\tilde{\gamma} = \gamma$. Clearly the tangent vector to the curve
$\tilde{\gamma}(t)$ is given by $K^\sharp \in
T^{\mathrm{h}}_{\tilde{\gamma}(t)}G$ and hence $\tilde{\gamma}(t)$ is the
horizontal lift of the curve $\gamma(t)$. Also note that
$\tilde{\gamma}(0) = L(p)$. As such we can identify the right
$H$-compensator $\tilde{h}_R^{(L)}(p,e^{tK})$ with $\tilde{h}(\gamma(t))$
of \refe{eqn:localconnectionproperty}. We can thus write
\begin{equation}
\omega^{(L)}(\pi_*( K^\sharp\big|_{L(p)})) = -
\frac{\mathrm{d}}{\mathrm{d}t} \tilde{h}_R^{(L)}(p,e^{tK})\Big|_{t=0}
\end{equation}
and hence
\begin{equation}
W^{(L)}_R (K) = \omega^{(L)}(\pi_*( K^\sharp\big|_{L(p)})) \quad
\mathrm{for} \quad K \in \mathfrak{k}.
\end{equation}
On the other hand consider $W^{(L)}_R(A)$ for $A \in \mathfrak{h}$. We
have
\begin{equation}
R_{h_1}^{(L)} (L(p)h) = L(p) h_1 h = L(r_{h_1}^{(L)}(p))
\tilde{h}_{R}^{(L)}(p,h_1)h \nonumber
\end{equation}
and using the fact that $r_{h}^{(L)}(p) = p$ we thus find
\begin{subequations}
\begin{equation}
\tilde{h}_{R}^{(L)}(p,e^{tH}) = e^{tH}
\end{equation}
and hence
\begin{equation}
W^{(L)}_R(H) = -\frac{\mathrm{d}}{\mathrm{d}t}
\tilde{h}_R^{(L)}(p,e^{tH})\Big|_{t=0} = -H.
\end{equation}
\end{subequations}
For the ease of notation we shall for the rest of this section drop the
explicit $L(p)$ dependence on $\pi_* A^\sharp$.

Now consider the infinitesimal versions of
\refec{eqn:comprg's}{eqn:comphr's}. We find with $g_1 = e^{\epsilon_1 A}$
and $g_2 = e^{\epsilon_2 B}$
\begin{subequations}
\begin{align}
\label{eqn:rginf} [\pi_* A^\sharp, \pi_* B^\sharp] & = \pi_* [A,B]^\sharp
+
\pi_*[W^{(L)}_R(A),B]^\sharp - \pi_*[W^{(L)}_R(B),A]^\sharp \\
W^{(L)}_R([A,B]) & = [W^{(L)}_R (A),W^{(L)}_R (B)] + \pi_* A^\sharp
W^{(L)}_R (B) - \pi_* B^\sharp
W^{(L)}_R (A) \nonumber \\
\label{eqn:hrinf} & \hspace{12pt} + {} W^{(L)}_R\left(
[W^{(L)}_R(B),A]\right) - W^{(L)}_R\left([W^{(L)}_R(A),B] \right).
\end{align}
\end{subequations}
Now, in the case where $A,B \in \mathfrak{k}$ we find that
\refe{eqn:rginf} corresponds to the expression for the torsion as given in
\refe{eqn:comptorsion}, whereas \refe{eqn:hrinf} corresponds to the
expression for the curvature as given in \refe{eqn:compcurvature}.

The infinitesimal versions of
\refec{eqn:gaugetransformrh}{eqn:gaugetransformrightHcomp} read
\begin{subequations}
\begin{align}
E_A^{(Lh)} & = \Lambda_{A}{}^B(h^{-1}) E_B^{(L)} \\
\label{eqn:spincongaugetrafo} W^{(Lh)}_R(\mathrm{Ad}_{h(p)^{-1}}A) & =
h(p)^{-1} W^{(L)}_R(A)h(p) + h(p)^{-1} \pi_* A^\sharp h(p).
\end{align}
\end{subequations}
In the case where $A \in \mathfrak{k}$ we find using
$\pi_*((\mathrm{Ad}_{h(p)^{-1}}A)^\sharp\big|_{L(p)h(p)}) =
\pi_*(A^\sharp\big|_{L(p)})$, see \refe{eqn:adhpi}, that
\refe{eqn:spincongaugetrafo} gives the transformation of the spin
connection under gauge transformations, cf.\
\refe{eqn:gaugetransformconnection}.

\subsubsection{Composition rules of mixed left and right actions}
Finally let us consider the successive action of $L_g^{(L)}$ and
$R_g^{(L)}$ on $L(p)$. Proceeding in the same way as when deriving the
composition rules for the left $H$-compensators,
\refec{eqn:complg's}{eqn:comprulelefth}, we find
\begin{subequations}
\begin{align}
\label{eqn:leftrightcomprulemaps} l_{g_1} \circ r_{g_2}^{(L)}(p)& =
r^{(L)}_{\mathrm{Ad}_{\tilde{h}_L^{(L)}(p,g_1)}g_2} \circ l_{g_1}(p)\\
\label{eqn:leftrightcomprule}
\tilde{h}_R^{(L)}(l_{g_1}(p),\mathrm{Ad}_{\tilde{h}_L^{(L)}(p,g_1)}g_2) &
=
\tilde{h}_L^{(L)}(r^{(L)}_{g_2}(p),g_1)\tilde{h}_R^{(L)}(p,g_2)\tilde{h}_L^{(L)}(p,g_1)^{-1}.
\end{align}
\end{subequations}
The infinitesimal versions of \refe{eqn:leftrightcomprulemaps} and
\refe{eqn:leftrightcomprule} with $g_1 = e^{\epsilon_1 A}$ and $g_2 =
e^{\epsilon_2 B}$ read
\begin{subequations}
\begin{align}
\label{eqn:leftrightcomprulemaps1} [\pi_* A^\flat, \pi_* B^\sharp] & = \pi_* [W^{(L)}_L(A), B]^\sharp\\
\label{eqn:leftrightcomprule1} [W^{(L)}_L(A),W^{(L)}_R(B)] -
\pi_*B^\sharp[W^{(L)}_L(A)] & = W^{(L)}_R([W^{(L)}_L(A),B]) - \pi_*
A^\flat[W^{(L)}_R(B)].
\end{align}
\end{subequations}

\section{Isometries}\label{sec:isometries}
When studying the geometry of superspace we do not, in general, have a
definition of a metric on the superspace \cite{bluebook,west}. Thus the
notion of isometry from ordinary geometry, \ie transformations which leave
the metric invariant, cannot be carried over to supergeometry. Instead we
must work with a definition of isometries based on the geometrical objects
at hand, that is the frame and connection. Imposing that the frame and
connection remain invariant under an isometry turns out to be too
restrictive and must be relaxed by demanding them to be invariant only up
to a gauge transformation, see e.g.\ \cite{hep-th/9812087}. This is
required to be a single gauge transformation transforming frame and
connection together, not independent transformations for each quantity.
For instance, the map $f:G/H\to G/H$ will be an isometry if
\begin{subequations}
\begin{alignat}{2}
\label{eqn:isome} f^\ast \left(E^A_{(L)}\big|_{f(p)}\right) &=
E^A_{(Lh)}\big|_p &&= E^B_{(L)}\big|_p \Lambda_B{}^A(h)\\
\label{eqn:isomomega} f^\ast \left(\omega^{(L)}\big|_{f(p)}\right) &=
\omega^{(Lh)}\big|_p &&= h^{-1} \omega^{(L)}\big|_p h + h^{-1}\extd h,
\end{alignat}
\end{subequations}
for some $h\in H$ which, as a gauge transformation, need not be constant.
It can be shown that if we impose such a condition on the map $f$ for one
choice of local section $L$ then it will automatically be satisfied for
other choices of $L$ with a different value for the gauge transformation
$h$.

As stated earlier the left action $L_g$ on $G$ induces a map $l_g$ on the
coset space $G/H$. Such maps can be thought of as isometries of the coset
space, as can be demonstrated by considering how the frame and connection
transform under the action of $l_g$. Recall from \refe{eqn:theequation}
that the pullback of the Maurer-Cartan form under a local section provides
us with both the local coframe and connection. Thus we will consider how
$L^\ast\zeta$ behaves under a pullback by $l_g$.

Consider a curve in the base $\gamma:[0,1]\to G/H$ with tangent vector $X$
and $\gamma(0)=p$. Using a matrix representation we have
\begin{equation}\label{eqn:pullbackzeta}
L^\ast\big(\zeta|_{L(p)}\big) \big(X|_p\big) = L(p)^{-1} \dd{t}
L(\gamma(t))\Big|_{t=0}.
\end{equation}
Now the curve $l_g\circ\gamma$ has tangent vector ${l_g}_\ast X$ and thus
\begin{align}
\nonumber L^\ast\big(\zeta|_{L(l_g(p))}\big) \big({l_g}_\ast (X|_p)
\big) &= L(l_g(p))^{-1} \dd{t} L(l_g(\gamma(t)))\Big|_{t=0} \\
\nonumber &=  L(l_g(p))^{-1} \dd{t} \left(g L(\gamma(t))
\tilde{h}_L^{(L)}(\gamma(t),g)^{-1} \right) \Big|_{t=0} \\
\nonumber &= \tilde{h}_L^{(L)}(p,g) L(p)^{-1}  \dd{t} \left( L(\gamma(t))
\tilde{h}_L^{(L)}(\gamma(t),g)^{-1} \right) \Big|_{t=0} \\
\nonumber &= \tilde{h}_L^{(L)}(p,g) L^\ast\big(\zeta|_{L(p)}\big)
\big(X|_p\big) \tilde{h}_L^{(L)}(p,g)^{-1}  \\ \nonumber
 &\hspace{1.5cm}  + \tilde{h}_L^{(L)}(p,g) \extd
\tilde{h}_L^{(L)}(p,g)^{-1} \big(X|_p\big).
\end{align}
Here we have used \refe{eqn:lgproperty} to obtain the second and third
lines, the final line follows from evaluating the derivative and using
\refe{eqn:pullbackzeta}. From this we thus see that
\begin{equation}
l_g^\ast L^\ast \big(\zeta|_{L(l_g(p))}\big) =
\mathrm{Ad}_{\tilde{h}_L^{(L)}(p,g)} L^\ast \big(\zeta|_{L(p)}\big) +
\tilde{h}_L^{(L)}(p,g) \extd \tilde{h}_L^{(L)}(p,g)^{-1}.
\end{equation}
If we now decompose this equation into its $\mathfrak{k}$ and
$\mathfrak{h}$ parts, we find
\begin{subequations}
\begin{align}
l_g^\ast\big( E^A_{(L)}|_{l_g(p)}\big) &= E^B_{(L)}|_p
\Lambda_B{}^A\big(\tilde{h}_L^{(L)}(p,g)^{-1}\big), \\
l_g^\ast \big(\omega^{(L)}|_{l_g(p)}\big) &=
\mathrm{Ad}_{\tilde{h}_L^{(L)}(p,g)} \omega^{(L)}|_p +
\tilde{h}_L^{(L)}(p,g) \extd \tilde{h}_L^{(L)}(p,g)^{-1}.
\end{align}
\end{subequations}
Comparing with \refec{eqn:isome}{eqn:isomomega} we see that the pulled
back coframe and connection from $l_g(p)$ to $p$ are simply a gauge
transform of the coframe and connection already at $p$; the parameter of
the gauge transformation is given by the $H$-compensator
$\tilde{h}_L^{(L)}(p,g)$.

From how the coframe transforms under the pullback by $l_g$ we may deduce
how the frame transforms under the pushforward by $l_g$. We find
\begin{equation}
{l_g}_\ast \big(E_A^{(L)}|_p\big) =
\Lambda_A{}^B\big(\tilde{h}_L^{(L)}(p,g)^{-1}\big) E_B^{(L)}|_{l_g(p)}.
\end{equation}

The maps $l_g$ therefore are isometries of $G/H$. They may be composed as
in \refe{eqn:complg's} and thus form a group of isometries isomorphic to
$G$. The maps $l_g$ may, however, not be all the isometries. For instance,
in the case when the normalizer $N(H)=\{g\in G : gHg^{-1}=H\}$ is
non-trivial then right multiplication in $G$ by an element $g\in N(H)$
results in a well defined map on the coset space which, if non-trivial,
i.e.\ $g\notin H$, also satisfies the conditions for it to be an isometry.
For a more detailed discussion of this see
\cite{lightblue,sym_coset_space}.

For an infinitesimal isometry given by $l_g$ with $g=e^{\epsilon A}$ we
have the associated supervector field $\pi_\ast A^\flat$, c.f.\
\refe{eqn:pistarAflatdefinition}. The supervectors $\pi_\ast A^\flat$ are
therefore Killing supervectors for the coset space, \ie supervectors which
give infinitesimal isometries. We see from \refe{eqn:algebrapiflats} that
the Killing supervectors satisfy an algebra. The set of $\pi_\ast
T_p^\flat$, $p=1,\ldots,\mathrm{dim}\,\mathfrak{g}$, form a set of
independent Killing supervectors, and from \refe{eqn:algebrapiflats}
satisfy
\begin{equation}
 [\pi_\ast T_p^\flat, \pi_\ast T_q^\flat] =  -f_{pq}{}^r \pi_\ast T_r^\flat.
\end{equation}
The definition of Killing supervectors will be discussed in more detail in
Section \ref{sec:killing}.

\section{Derivations on associated bundles}
\label{sec:derivations} In this section we shall consider derivations, --
such as the covariant derivative, the Lie derivative and the so-called
$H$-covariant Lie derivative -- on associated bundles. A derivation, or
more precisely a graded derivation, is here defined as a linear map on an
abstract algebra satisfying the graded Leibniz rule.

\subsection{Covariant derivative}
\label{sec:covder} In the following we shall define the covariant
derivative $\nabla_{\pi_*\left(K^\sharp|_{L(p)}\right)}$ on a local
section $s(p) = [(L(p),\xi(p))]$ of a general associated bundle in the
direction of the push-forward of an even horizontal vector $K^\sharp$ in
terms of the map $R_g^{(L)}$. Before we do this let us, however, first
review the standard definition of the covariant derivative on associated
bundles, see \cite{nakahara}.

Consider a curve $\gamma(t)$ in the base with $t \in [0,1]$ and $\gamma(0)
= p$. We can then write, using \refe{eqn:horizontallift},
\begin{align}
s(\gamma(t)) & = \Big\lbrack\Big(L(\gamma(t)),\xi(\gamma(t))\Big)\Big\rbrack \nonumber \\
& = \Big\lbrack\Big(\tilde{\gamma}(t)\tilde{h}(t)^{-1},\xi(\gamma(t))\Big)\Big\rbrack\nonumber\\
& = \Big\lbrack\Big(\tilde{\gamma}(t),
\rho(\tilde{h}(t)^{-1})\xi(\gamma(t))\Big)\Big\rbrack.
\end{align}
Now, setting $\eta(\gamma(t)) \equiv
\rho(\tilde{h}(t)^{-1})\xi(\gamma(t))$, we have for the standard
definition of the covariant derivative
\begin{equation}
\label{eqn:covdernakahara} \nabla_X s(p) \defined
\left\lbrack\left(\tilde{\gamma}(0),\frac{\mathrm{d}}{\mathrm{d}t}\eta(\gamma(t))
\Big|_{t=0}\right)\right\rbrack,
\end{equation}
where $X$ is the tangent vector to $\gamma(t)$ at $p$. From this we see
that a local section $s(\gamma(t))$ is parallel transported along
$\gamma(t)$ if $\eta$ is constant along $\gamma(t)$. It is easy to see
that the covariant derivative does not depend on the specific choice of
horizontal lift $\tilde{\gamma}(t)$.

On the other hand we shall now see that one can also define the covariant
derivative of a local section $s(p)$ in terms of the map $R_g^{(L)}$. We
will define
\begin{equation}
\label{eqn:covder} \nabla_{\pi_*\left(K^\sharp|_{L(p)}\right)} s(p)
\defined \lim_{\epsilon \to 0} \frac{1}{\epsilon} \left(
\left(\tilde{R}^{(L)}_{e^{\epsilon K}}\right)^{-1} s(r^{(L)}_{e^{\epsilon
K}}(p)) - s(p)\right).
\end{equation}
Using the definition of the map $\tilde{R}^{(L)}_g$ on $s(p)$, see
\refe{eqn:tildeRgL}, this can be rewritten as
\begin{align}
\nabla_{\pi_*\left(K^\sharp|_{L(p)}\right)} s(p) & = \lim_{\epsilon \to 0}
\frac{1}{\epsilon} \Big(\Big\lbrack\Big(\,(R_{e^{\epsilon
K}}^{(L)})^{-1}L(r^{(L)}_{e^{\epsilon K}}(p))\,,\,
\xi(r^{(L)}_{e^{\epsilon K}}(p))\,\Big)\Big\rbrack \nonumber \\
& \hspace{34pt} - {}
\Big\lbrack\Big(L(p),\xi(p)\Big)\Big\rbrack\Big) \nonumber \\
& = \lim_{\epsilon \to 0} \frac{1}{\epsilon}
\Big(\Big\lbrack\Big(\,R^{(L)}_{\mathrm{Ad}_{\tilde{h}_R^{(L)}(p,e^{\epsilon
K})}e^{-\epsilon K}}L(r^{(L)}_{e^{\epsilon K}}(p))\,,\,
\xi(r^{(L)}_{e^{\epsilon K}}(p))\,\Big)\Big\rbrack \nonumber \\
& \hspace{34pt} - {}\Big\lbrack\Big(L(p),\xi(p)\Big)\Big\rbrack\Big)
\nonumber \\
& = \lim_{\epsilon \to 0} \frac{1}{\epsilon}
\Big(\Big\lbrack\Big(\,L(p)\tilde{h}^{(L)}_R(p,e^{\epsilon
K})^{-1}\,,\,\xi(r^{(L)}_{e^{\epsilon K}}(p))\,\Big)\Big\rbrack \nonumber \\
& \hspace{34pt} - {} \Big\lbrack\Big(L(p),\xi(p)\Big)\Big\rbrack\Big)
\nonumber \\
& = \lim_{\epsilon \to 0} \frac{1}{\epsilon}
\left(\Big\lbrack\Big(\,L(p)\,,\, \rho(\tilde{h}^{(L)}_R(p,e^{\epsilon
K})^{-1})\xi(r^{(L)}_{e^{\epsilon
K}}(p)) - \xi(p)\,\Big)\Big\rbrack\right) \nonumber \\
\label{eqn:covversusnakaharacov}& = \Big\lbrack\Big(L(p),
\frac{\mathrm{d}}{\mathrm{d}t}\left(\rho(\tilde{h}^{(L)}_R(p,e^{t
K})^{-1}) \xi(r^{(L)}_{e^{t K}}(p))\right)\Big|_{t = 0}\Big)\Big\rbrack.
\end{align}
Note that in deriving this we have used
\begin{displaymath}
(R_g^{(L)})^{-1} (L(p)h) =
R^{(L)}_{\mathrm{Ad}_{\tilde{h}_R^{(L)}(p,g)}g^{-1}} (L(p)h)
\end{displaymath}
as well as the equivalence relation \refe{eqn:equivrelvector}. Noting that
$\tilde{\gamma}(0) = L(p)$ and $\tilde{h} = \tilde{h}_R^{(L)}$ we easily
see from \refe{eqn:covversusnakaharacov} that our definition of the
covariant derivative, \refe{eqn:covder}, is -- in the specific case of the
vector $\pi_* \left(K^\sharp|_{L(p)}\right)$ -- equivalent to the standard
definition given by \refe{eqn:covdernakahara}. Note that although the
range of the map $R^{(L)}_{e^{\epsilon K}}$ is considered to be restricted
to $\pi^{-1}(U)$ this does not pose a problem for our definition of the
covariant derivative as $\epsilon$ can always be chosen sufficiently small
such that $R^{(L)}_{e^{\epsilon K}} L(p) \in \pi^{-1}(U)$.

Now, from \refe{eqn:covversusnakaharacov} we easily find
\begin{align}
\nabla_{\pi_*\left(K^\sharp|_{L(p)}\right)} s(p) & = \Big\lbrack
\Big(\,L(p)\,,\, \frac{\mathrm{d}}{\mathrm{d}t}\xi(r^{(L)}_{e^{t K}}(p))
\Big|_{t=0} + \frac{\mathrm{d}}{\mathrm{d}t}\rho(\tilde{h}^{(L)}_R(p,e^{t
K})^{-1})\Big|_{t = 0} \xi(p)
\,\Big) \Big\rbrack \nonumber \\
\label{eqn:covdercomp} & = \Big\lbrack \Big(\,L(p)\,,\, \pi_*
(K^\sharp|_{L(p)}) [\xi(p)] +
\rho\big(\omega^{(L)}(\pi_*(K^\sharp|_{L(p)}))\big) \xi(p)\,\Big)
\Big\rbrack.
\end{align}
Here we consider $ \omega^{(L)}$ in the representation appropriate for
acting on $\xi(p)$. It is easy to see that the covariant derivative is
invariant under the equivalence transformations on associated bundles, see
\refe{eqn:equivrel}. We have
\begin{align}
\nabla_{\pi_*\left(K^\sharp|_{L(p)}\right)} [(L(p)h, \rho(h^{-1})\xi(p))]&
\nonumber \\ & \hspace{-80pt} = \lim_{\epsilon \to 0} \frac{1}{\epsilon}
\Big(\Big\lbrack\Big(\,(R_{e^{\epsilon
K}}^{(L)})^{-1}\big(L(r^{(L)}_{e^{\epsilon K}}(p))h\big)\,,\,
\rho(h^{-1})\xi(r^{(L)}_{e^{\epsilon K}}(p))\,\Big)\Big\rbrack \nonumber \\
& \hspace{-46pt} - {}
\Big\lbrack\Big(L(p)h,\rho(h^{-1})\xi(p)\Big)\Big\rbrack\Big) \nonumber
\\ & \hspace{-80pt} = \lim_{\epsilon \to 0} \frac{1}{\epsilon}
\Big(\Big\lbrack\Big(\,L(p)\tilde{h}^{(L)}_R(p,e^{\epsilon
K})^{-1}h\,,\,\rho(h^{-1})\xi(r^{(L)}_{e^{\epsilon K}}(p))\,\Big)\Big\rbrack \nonumber \\
& \hspace{-46pt} - {}
\Big\lbrack\Big(L(p)h,\rho(h^{-1})\xi(p)\Big)\Big\rbrack\Big) \nonumber \\
\nonumber & \hspace{-80pt} = \nabla_{\pi_*\left(K^\sharp|_{L(p)}\right)}
[(L(p),\xi(p))].
\end{align}
The covariant derivative of $s(p)$ is therefore, as a section, well
defined.

Using $\pi_* \left(K^\sharp|_{L(p)}\right) = X^A E_A^{(L)}$ and
$\nabla_{E_A^{(L)}} \equiv \nabla_A^{(L)}$, we have
\begin{equation}
\label{eqn:covdercomp1} \nabla_A^{(L)} s(p) = \Big\lbrack \Big(\,L(p)\,,\,
E_A^{(L)} [\xi(p)] + \rho\big(\omega^{(L)}_A\big) \xi(p)\,\Big)
\Big\rbrack,
\end{equation}
which will transform under a change of local section as
\begin{equation}
\nabla_A^{(L)} s(p) = \Lambda_A{}^B(h) \nabla_B^{(Lh)}s(p).
\end{equation}

From \refe{eqn:covdercomp1} we see that our definition of the covariant
derivative, \refe{eqn:covder}, gives us an expression for the covariant
derivative in the direction of the basis vectors $E_A^{(L)}$. As such it
extends to a derivative in the direction of an arbitrary vector field $X =
X^A E_A^{(L)}$, although our initial definition was only given for vector
fields $X = X^A E_A^{(L)}$ with $X^A$ constant. Note that for convenience
of notation we shall for the rest of this section drop the explicit $L(p)$
dependence on the vectors $\pi_*(K^\sharp\big|_{L(p)})$ as well as on
$E_A^{(L)}$.

As a practical example let us consider calculating the covariant
derivative of a local section of a contravariant vector bundle $X(p) =
[(L(p),\{X^A\})] = X^A E_A$. We have
\begin{align}
\nabla_{\pi_* K^\sharp} X(p) & =  \lim_{\epsilon \to 0} \frac{1}{\epsilon}
\Big(\Big\lbrack\Big(\,L(p)\tilde{h}^{(L)}_R(p,e^{\epsilon K})^{-1}\,,\,
\Big\{X^A(r^{(L)}_{e^{\epsilon
K}}(p))\Big\}\,\Big)\Big\rbrack \nonumber \\
& \hspace{34pt} -{} \Big\lbrack\Big(L(p),\{X^A(p)\}\Big)\Big\rbrack\Big) \nonumber \\
& = \lim_{\epsilon \to 0} \frac{1}{\epsilon}
\Big(\Big\lbrack\Big(\,L(p)\,,\, \Big\{X^A(r^{(L)}_{e^{\epsilon
K}}(p))\Lambda_A{}^B(\tilde{h}_R^{(L)}(p,e^{\epsilon K}))\Big\}\,\Big)\Big\rbrack \nonumber \\
& \hspace{34pt} -{} \Big\lbrack\Big(L(p),\{X^A(p)\}\Big)\Big\rbrack\Big),
\label{eqn:covdervector1}
\end{align}
where we have used the equivalence relation for tensor bundles, see
\refec{eqn:equivreltensor1}{eqn:equivreltensor2}. Using
\refec{eqn:expansionhr}{eqn:infinitesimalcoadjoint} we can expand
$\Lambda_A{}^B(\tilde{h}_R^{(L)}(p,e^{\epsilon K}))$ as
\begin{equation}
\Lambda_A{}^B(\tilde{h}_R^{(L)}(p,e^{\epsilon K})) = \delta_A{}^B +
\epsilon \omega^{I}(\pi_* K^\sharp)f_{IA}{}^B +
\mathcal{O}(\epsilon^2)\nonumber,
\end{equation}
where we have also dropped the $L(p)$ dependence on $\omega$. Using
\refe{eqn:expansionf(r)} we can expand $X^A(r^{(L)}_{e^{\epsilon K}}(p))$
as
\begin{displaymath}
X^A(r^{(L)}_{e^{\epsilon K}}(p)) = X^A(p) + \epsilon \pi_* K^\sharp
[X^A(p)] + \mathcal{O}(\epsilon^2),
\end{displaymath}
which allows us to rewrite \refe{eqn:covdervector1} as
\begin{equation}
\label{eqn:covder_vector} \nabla_{\pi_* K^\sharp} X(p) =
\Big\lbrack\Big(L(p),\Big\{\pi_*K^\sharp[X^A(p)] + X^B(p)
\omega^{I}(\pi_*K^\sharp)f_{IB}{}^A\Big\}\Big)\Big\rbrack.
\end{equation}

Finally let us consider the commutator of two covariant derivatives
\begin{multline*}
[\nabla_{\pi_*K_1^\sharp},\nabla_{\pi_*K_2^\sharp}] s(p) =
\lim_{\stackrel{\scriptstyle \epsilon_1 \to 0}{\scriptstyle \epsilon_2 \to
0}} \frac{1}{\epsilon_1 \epsilon_2}\Big( \Big\lbrack\Big( L(p), \\
\hspace{10pt} \rho(\tilde{h}_R^{(L)}(p,e^{\epsilon_1 K_1})^{-1})
\rho(\tilde{h}_R^{(L)}(r_{e^{\epsilon_1K_1}}(p),e^{\epsilon_2 K_2})^{-1})
\xi(r_{e^{\epsilon_2K_2}}(r_{e^{\epsilon_1K_1}}(p))) \\
{}-\rho(\tilde{h}_R^{(L)}(p,e^{\epsilon_2 K_2})^{-1})
\rho(\tilde{h}_R^{(L)}(r_{e^{\epsilon_2K_2}}(p),e^{\epsilon_1 K_1})^{-1})
\xi(r_{e^{\epsilon_1K_1}}(r_{e^{\epsilon_2K_2}}(p)))
\Big)\Big\rbrack\Big).
\end{multline*}
Expanding and taking the limit we have
\begin{multline*}
[\nabla_{\pi_*K_1^\sharp},\nabla_{\pi_*K_2^\sharp}] s(p) = \Big\lbrack
\Big(L(p),\, [\omega(\pi_* K_1^\sharp),\omega(\pi_* K_2^\sharp)]\xi(p)
\\  {}+ [\pi_* K_1^\sharp,\pi_* K_2^\sharp]\xi(p) +
\pi_*K_1^\sharp[\omega(\pi_* K_2^\sharp)]\xi(p) -
\pi_*K_2^\sharp[\omega(\pi_* K_1^\sharp)]\xi(p) \Big) \Big\rbrack.
\end{multline*}
Note that for the ease of notation we have suppressed the representation
$\rho$ on $\omega(\pi_* K_2^\sharp)$. Now, using the expressions for the
curvature as given in \refe{eqn:RXY}, we can rewrite this last equation as
\begin{multline*}
[\nabla_{\pi_*K_1^\sharp},\nabla_{\pi_*K_2^\sharp}] s(p)  = \Big\lbrack
\Big(L(p),
\\
R(\pi_* K_1^\sharp, \pi_* K_2^\sharp)\xi(p) + [\pi_* K_1^\sharp,\pi_*
K_2^\sharp]\xi(p) + \omega([\pi_* K_1^\sharp,\pi_*
K_2^\sharp])\xi(p))\Big) \Big\rbrack
\end{multline*}
and hence we finally have, using the expression for the covariant
derivative as given in \refe{eqn:covdercomp},
\begin{equation}
\label{eqn:commcovdev0}
[\nabla_{\pi_*K_1^\sharp},\nabla_{\pi_*K_2^\sharp}] s(p)  = R(\pi_*
K_1^\sharp, \pi_* K_2^\sharp)s(p) + \nabla_{[\pi_* K_1^\sharp,\pi_*
K_2^\sharp ]}s(p),
\end{equation}
where $R(\pi_* K_1^\sharp, \pi_* K_2^\sharp)s(p)$ is an abbreviation for
the action of $\tilde{R}_H^{(L)}$ on the section $s(p)$ with $H = R(\pi_*
K_1^\sharp, \pi_* K_2^\sharp)$. Now, setting $\pi_*K_1^\sharp = X^A E_A$
and $\pi_*K_2^\sharp = Y^A E_A$ we can rewrite the left hand side of this
equation as
\begin{align}
[\nabla_{X^A E_A},\nabla_{Y^B E_B}] s(p) & = X^A (\nabla_A Y)^B
\nabla_B s(p) - Y^B (\nabla_B X)^A \nabla_A s(p) \nonumber \\
& \hspace{15pt} + Y^B X^A [\nabla_A, \nabla_B] s(p)\nonumber \\
& = Y^B X^A [\nabla_A,\nabla_B] s(p)  \nonumber \\
& \hspace{15pt} - {} Y^B X^A \left(\omega_{AB}{}^C \nabla_C + (-1)^{AB}
\omega_{BA}{}^C \nabla_C\right) s(p) \nonumber,
\end{align}
where we have used \refe{eqn:covder_vector} for constant $X$. The right
hand side of \refe{eqn:commcovdev0} can be rewritten as
\begin{equation}
R(X^A E_A, Y^B E_B) + \nabla_{[X^A E_A,Y^B E_B]} = Y^B X^A\left( R_{AB} +
\mathcal{C}_{AB}{}^C \nabla_C \right) \nonumber
\end{equation}
and we hence find in total, using the component expression of the torsion,
\refe{eqn:comptorsion},
\begin{equation}
[\nabla_A, \nabla_B]  = R_{AB}  + T_{AB}{}^C \nabla_C .
\end{equation}

\subsection{Lie derivative}
In this section we will introduce the Lie derivative on a local section
$s(p)$ in the direction of an isometry. We define
\begin{equation}
\mathcal{L}_{\pi_* A^\flat} s(p) \defined \lim_{\epsilon \to 0}
\frac{1}{\epsilon} \left( \left(\tilde{L}_{e^{\epsilon A}}\right)^{-1}
s(l_{e^{\epsilon A}}(p)) - s(p)\right).
\end{equation}
We can rewrite the Lie derivative as
\begin{align}
\mathcal{L}_{\pi_* A^\flat} s(p) & = \lim_{\epsilon \to 0}
\frac{1}{\epsilon} \left( \Big\lbrack\Big( L_{e^{-\epsilon A}}
\left(L(l_{e^{\epsilon A}}(p))\right),\xi\left(l_{e^{\epsilon
A}}(p)\right) \Big)\Big\rbrack -
\Big\lbrack\Big(L(p) , \xi(p) \Big)\Big\rbrack \right) \nonumber \\
& = \lim_{\epsilon \to 0} \frac{1}{\epsilon} \left( \Big\lbrack\Big(
e^{-\epsilon A} L(l_{e^{\epsilon A}}(p)),\xi\left(l_{e^{\epsilon
A}}(p)\right) \Big)\Big\rbrack - \Big\lbrack\Big(L(p) , \xi(p)
\Big)\Big\rbrack \right) \nonumber \\
& = \lim_{\epsilon \to 0} \frac{1}{\epsilon} \left( \Big\lbrack\Big(
L(p)\tilde{h}_L^{(L)}(p,e^{-\epsilon A}),\xi\left(l_{e^{\epsilon
A}}(p)\right) \Big)\Big\rbrack - \Big\lbrack\Big(L(p) , \xi(p)
\Big)\Big\rbrack \right) \nonumber \\
& = \lim_{\epsilon \to 0} \frac{1}{\epsilon} \Big\lbrack\Big( L(p),
\rho(\tilde{h}_L^{(L)}(p,e^{-\epsilon A}))\xi\left(l_{e^{\epsilon
A}}(p)\right) - \xi(p) \Big)\Big\rbrack \nonumber \\
\label{eqn:liederddt}& = \Big\lbrack\Big(L(p),
\frac{\mathrm{d}}{\mathrm{d}t}\left(\rho(\tilde{h}^{(L)}_L(p,e^{-t A}))
\xi(l_{e^{t A}}(p))\right)\Big|_{t = 0}\Big)\Big\rbrack.
\end{align}
One should note the similarity to the expression of the covariant
derivative given in \refe{eqn:covversusnakaharacov}. Note also that,
although the range of the map $L_{e^{\epsilon A}}$ is considered to be
restricted to $\pi^{-1}(U)$, this does not pose a problem for our
definition of the Lie derivative as $\epsilon$ can always be chosen
sufficiently small such that $L_{e^{\epsilon A}} L(p) \in \pi^{-1}(U)$.
Now we easily find
\begin{align}
\mathcal{L}_{\pi_* A^\flat} s(p) & = \Big\lbrack \Big(\,L(p)\,,\,
\frac{\mathrm{d}}{\mathrm{d}t}\xi(l_{e^{t A}}(p)) \Big|_{t=0} +
\frac{\mathrm{d}}{\mathrm{d}t}\rho(\tilde{h}^{(L)}_L(p,e^{-t A}))\Big|_{t
= 0} \xi(p)
\,\Big) \Big\rbrack \nonumber \\
\label{eqn:liecomp} & = \Big\lbrack \Big(\,L(p)\,,\, \pi_* A^\flat
[\xi(p)] + \rho\big(W^{(L)}_L(A)\big) \xi(p)\,\Big) \Big\rbrack,
\end{align}
where we have used \refec{eqn:pistarAflatdefinition}{eqn:defW}. Here we
consider $ W^{(L)}_L$ to be in the representation appropriate for acting
on $\xi(p)$. It is easy to see that the Lie derivative is invariant under
the equivalence transformations $[(L(p), \xi(p))] = [(L(p)h,
\rho(h^{-1})\xi(p))]$. The Lie derivative of $s(p)$ is therefore, as a
section, well defined. Again for the ease of notation we shall for the
rest of this section drop the explicit $L(p)$ on the $E_A^{(L)}$.

As an example we shall now calculate the Lie derivative of a vector and a
one-form, respectively. First consider a local section of a vector bundle
$X(p) = [(L(p),\{X^A\})] = X^A E_A$. We have
\begin{align}
\mathcal{L}_{\pi_* A^\flat} X(p) & =  \lim_{\epsilon \to 0}
\frac{1}{\epsilon}
\Big(\Big\lbrack\Big(\,L(p)\tilde{h}^{(L)}_L(p,e^{-\epsilon A})\,,\,
\Big\{X^A(l_{e^{\epsilon
A}}(p))\Big\}\,\Big)\Big\rbrack \nonumber \\
& \hspace{34pt} -{} \Big\lbrack\Big(L(p),\{X^A(p)\}\Big)\Big\rbrack\Big) \nonumber \\
& = \lim_{\epsilon \to 0} \frac{1}{\epsilon}
\Big(\Big\lbrack\Big(\,L(p)\,,\, \Big\{X^A(l_{e^{\epsilon
A}}(p))\Lambda_A{}^B(\tilde{h}_L^{(L)}(p,e^{-\epsilon
A})^{-1})\Big\}\,\Big)\Big\rbrack \nonumber \\ \label{eqn:lie_vector1} &
\hspace{34pt} -{} \Big\lbrack\Big(L(p),\{X^A(p)\}\Big)\Big\rbrack\Big),
\end{align}
where we have used the equivalence relation for tensor bundles, see
\refec{eqn:equivreltensor1}{eqn:equivreltensor2}. Using
\refec{eqn:expansionhl}{eqn:infinitesimalcoadjoint} we can expand
$\Lambda_A{}^B(\tilde{h}^{(L)}_L(p,e^{-\epsilon A})^{-1})$ as
\begin{equation}
\Lambda_A{}^B(\tilde{h}^{(L)}_L(p,e^{-\epsilon A})^{-1}) = \delta_A{}^B
+\epsilon W_L^{I}f_{IA}{}^B + \mathcal{O}(\epsilon^2)\nonumber,
\end{equation}
where we have now also dropped the $L(p)$ dependence on $W_L$. Using
\refe{eqn:pistarAflatdefinition} we can expand $X^A(l_{e^{\epsilon
A}}(p))$ as
\begin{displaymath}
X^A(l_{e^{\epsilon A}}(p)) = X^A(p) + \epsilon \pi_* A^\flat [X^A(p)] +
\mathcal{O}(\epsilon^2),
\end{displaymath}
which allows us to rewrite \refe{eqn:lie_vector1} as
\begin{equation}
\label{eqn:lie_vector} \mathcal{L}_{\pi_* A^\flat} X(p) =
\Big\lbrack\Big(L(p),\Big\{\pi_*A^\flat[X^A(p)] + X^B(p)
W_L^{I}(A)f_{IB}{}^A\Big\}\Big)\Big\rbrack.
\end{equation}
In the particular case of the Lie derivative of the basis vector $E_A$ we
find from this
\begin{align}
\mathcal{L}_{\pi_* A^\flat} E_A & =
\Big\lbrack\Big(L(p),\Big\{\delta_A{}^B
W_L^{I}(A)f_{IB}{}^C\Big\}\Big)\Big\rbrack \nonumber \\
& = \Big\lbrack\Big(L(p),\Big\{ W_L^{I}(A)f_{IA}{}^B
\delta_B{}^C\Big\}\Big)\Big\rbrack \nonumber \\
& = W_L^{I}(A)f_{IA}{}^B E_B.
\end{align}

By a similar calculation as in the case of the Lie derivative of a vector
we find for the Lie derivative of a one-form $\phi(p) =[(L(p),\{\phi_A\})]
= E^A \phi_A $
\begin{align}
\label{eqn:lie_form} \mathcal{L}_{\pi_* A^\flat} \phi(p) =
\Big\lbrack\Big(L(p),\Big\{\pi_*A^\flat[\phi_A(p)] - W_L^{I}(A)f_{IA}{}^B
\phi_B(p) \Big\}\Big)\Big\rbrack
\end{align}
and hence in the particular case of the Lie derivative of the basis vector
$E^A$
\begin{align}
\mathcal{L}_{\pi_* A^\flat} E^A & = \Big\lbrack\Big(L(p),\Big\{-
W_L^{I}(A)f_{IB}{}^C \delta_C{}^A(p) \Big\}\Big)\Big\rbrack \nonumber \\
& = \Big\lbrack\Big(L(p),\Big\{ - \delta_B{}^C W_L^{I}(A)f_{IC}{}^A
\Big\}\Big)\Big\rbrack \nonumber \\
& = - E^C W_L^{I}(A)f_{IC}{}^A.
\end{align}
Now, in the case of the connection one-from $\omega$ we find from
\refe{eqn:lie_form} together with the composition rule of the right and
left $H$-compensators, see \refe{eqn:leftrightcomprule1},
\begin{align}
\mathcal{L}_{\pi_* A^\flat} \omega(p) & =
\Big\lbrack\Big(L(p),\Big\{\pi_*A^\flat[\omega_A(p)] -
W_L^{I}(A)f_{IA}{}^B \omega_B(p) \Big\}\Big)\Big\rbrack \nonumber \\
& = \Big\lbrack\Big(L(p),\Big\{ E_A [\rho\big(W_L(A)\big)] -
[\rho\big(W_L(A)\big), \omega_A] \Big\}\Big)\Big\rbrack.
\end{align}

Now consider the algebra of two Lie derivatives. We find, writing for
simplicity $W_L(A)$ instead of $\rho\big(W_L(A)\big)$,
\begin{multline*}
[\mathcal{L}_{\pi_* A^\flat}, \mathcal{L}_{\pi_* B^\flat}] s(p) =
\Big\lbrack\Big(L(p)\,\,,\,\, [\pi_* A^\flat, \pi_* B^\flat]\xi(p) \\ + {}
(\pi_* A^\flat [W_L(B)] - \pi_* B^\flat [W_L(A)] + [W_L(A),W_L(B)])
\xi(p)\Big)\Big\rbrack
\end{multline*}
and hence, using \refe{eqn:intcondW1}, we find
\begin{align}
[\mathcal{L}_{\pi_* A^\flat}, \mathcal{L}_{\pi_* B^\flat}] s(p) & =
\Big\lbrack\Big(L(p)\,\,,\,\, [\pi_* A^\flat, \pi_* B^\flat]\xi(p)  +
W_L([B,A])\xi(p)\Big)\Big\rbrack.
\end{align}
Now, noting that the algebra element corresponding to the vector $[\pi_*
A^\flat, \pi_* B^\flat] = \pi_* [B,A]^\flat$ is given by $[B,A]$, cf.\
\refe{eqn:algebrapiflats}, we find
\begin{equation}
[\mathcal{L}_{\pi_* A^\flat}, \mathcal{L}_{\pi_* B^\flat}]  =
\mathcal{L}_{[\pi_* A^\flat, \pi_* B^\flat]}.
\end{equation}

Finally let us consider the commutator of the Lie derivative with the
covariant derivative on a local section $s(p)=[(L(p), \xi(p))]$. We find
\begin{align}
[\mathcal{L}_{\pi_* A^\flat}, \nabla_{\pi_* B^\sharp}] s(p) & =
\Big\lbrack\Big(L(p)\,, \,\, [\pi_* A^\flat, \pi_* B^\sharp] \xi(p) +
\pi_* A^\flat [\omega(\pi_*
B^\sharp)]\xi(p) \nonumber \\
& \hspace{55pt} - {} \pi_* B^\sharp[W_L(A)]\xi(p) +  [W_L(A), \omega (\pi_* B^\sharp)]\xi(p)\Big)\Big\rbrack \nonumber \\
& = \Big\lbrack\Big( L(p)\,,\,\, [\pi_* A^\flat, \pi_* B^\sharp] \xi(p) +
\omega([\pi_* A^\flat, \pi_* B^\sharp])\xi(p) \Big)\Big\rbrack \nonumber\\
& = \nabla_{[\pi_* A^\flat, \pi_* B^\sharp]} s(p), \nonumber
\end{align}
where we have used
\refec{eqn:leftrightcomprulemaps1}{eqn:leftrightcomprule1}. We thus have
\begin{equation}
\label{eqn:commutatorgradlie} [\mathcal{L}_{\pi_* A^\flat}, \nabla_{\pi_*
B^\sharp}] s(p) = \nabla_{(\mathcal{L}_{\pi_* A^\flat} \pi_* B^\sharp)}
s(p).
\end{equation}
Now, setting $\pi_* B^\sharp = X^A E_A$, $X^A$ constant, we have,
evaluating the left hand side of \refe{eqn:commutatorgradlie} directly
\begin{displaymath}
[\mathcal{L}_{\pi_* A^\flat}, \nabla_{\pi_* B^\sharp}] s(p) =
(\mathcal{L}_{\pi_* A^\flat} \pi_* B^\sharp)^A \nabla_{A}s(p) +
X^A[\mathcal{L}_{\pi_* A^\flat}, \nabla_{A}] s(p).
\end{displaymath}
Combining this with \refe{eqn:commutatorgradlie} we finally find
\begin{equation}
\label{eqn:commliecov0} [\mathcal{L}_{\pi_* A^\flat}, \nabla_{A}]  = 0.
\end{equation}

\subsection{$H$-covariant Lie derivative}
In this section we shall define the so-called $H$-covariant Lie
derivative, the characteristic property of which is that when
differentiating the frame it gives zero.

We define the $H$-covariant Lie derivative of a section $s(p)$ of a
general tensor bundle in the direction of an isometry as
\begin{equation}
\label{eqn:hcovdef} \mathbbm{L}^{(L)}_{\pi_* A^\flat} s(p)
\defined \lim_{\epsilon \to 0} \frac{1}{\epsilon} \left(
\left(\tilde{L}^{(L)}_{e^{\epsilon A}}\right)^{-1} s(l_{e^{\epsilon
A}}(p)) - s(p)\right).
\end{equation}
We can rewrite the $H$-covariant Lie derivative as
\begin{align}
\mathbbm{L}^{(L)}_{\pi_* A^\flat} s(p) & = \lim_{\epsilon \to 0}
\frac{1}{\epsilon} \left( \Big\lbrack\Big( L^{(L)}_{e^{-\epsilon A}}
\left(L(l_{e^{\epsilon A}}(p))\right),\xi\left(l_{e^{\epsilon
A}}(p)\right) \Big)\Big\rbrack -
\Big\lbrack\Big(L(p) , \xi(p) \Big)\Big\rbrack \right) \nonumber \\
& = \lim_{\epsilon \to 0} \frac{1}{\epsilon} \Big( \Big\lbrack\Big(
e^{-\epsilon A} L(l_{e^{\epsilon A}}(p))\tilde{h}^{(L)}_L(l_{e^{\epsilon
A}}, e^{-\epsilon A})^{-1},\xi\left(l_{e^{\epsilon A}}(p)\right)
\Big)\Big\rbrack \nonumber \\ & \hspace{34pt}- {}\Big\lbrack\Big(L(p) ,
\xi(p)
\Big)\Big\rbrack \Big) \nonumber \\
& = \lim_{\epsilon \to 0} \frac{1}{\epsilon} \Big( \Big\lbrack\Big(
e^{-\epsilon A} L(l_{e^{\epsilon A}}(p))\tilde{h}_L^{(L)}(p,e^{\epsilon
A}),\xi\left(l_{e^{\epsilon A}}(p)\right) \Big)\Big\rbrack \nonumber \\ &
\hspace{34pt}- {} \Big\lbrack\Big(L(p) , \xi(p)
\Big)\Big\rbrack \Big) \nonumber \\
& = \lim_{\epsilon \to 0} \frac{1}{\epsilon} \Big\lbrack\Big( L(p),
\xi\left(l_{e^{\epsilon
A}}(p)\right) - \xi(p) \Big)\Big\rbrack \nonumber \\
\label{eqn:hliederddt}& = \Big\lbrack\Big(L(p),
\frac{\mathrm{d}}{\mathrm{d}t}\xi(l_{e^{t A}}(p))\Big|_{t =
0}\Big)\Big\rbrack,
\end{align}
where we have used \refe{eqn:tildehlrelinverse}. Again note that, although
the range of the map $L^{(L)}_{e^{\epsilon A}}$ is considered to be
restricted to $\pi^{-1}(U)$, this does not pose a problem for our
definition of the $H$-covariant Lie derivative as $\epsilon$ can always be
chosen sufficiently small such that $L^{(L)}_{e^{\epsilon A}} L(p) \in
\pi^{-1}(U)$. Now we immediately find from \refe{eqn:hliederddt}
\begin{equation}
\label{eqn:hcovpi*} \mathbbm{L}^{(L)}_{\pi_* A^\flat} s(p)  = \Big\lbrack
\Big(\,L(p)\,,\, \pi_* A^\flat [\xi(p)]\,\Big) \Big\rbrack.
\end{equation}
It is easy to see that the $H$-covariant Lie derivative is invariant under
the equivalence transformations on associated bundles, see
\refe{eqn:equivrel}. We have
\begin{align}
\mathbbm{L}^{(L)}_{\pi_*A^\flat} [(L(p)h, \rho(h^{-1})\xi(p))] & \nonumber
\\ & \hspace{-70pt} = \lim_{\epsilon \to 0} \frac{1}{\epsilon} \Big(
\Big\lbrack\Big( L^{(L)}_{e^{-\epsilon A}} \left(L(l_{e^{\epsilon
A}}(p))h\right) ,\rho(h^{-1})\xi\left(l_{e^{\epsilon A}}(p)\right)
\Big)\Big\rbrack \nonumber \\ & \hspace{-70pt} \hspace{34pt} - {}
\Big\lbrack\Big(L(p)h,
\rho(h^{-1})\xi(p) \Big)\Big\rbrack \Big) \nonumber \\
& \hspace{-70pt} = \lim_{\epsilon \to 0} \frac{1}{\epsilon} \Big(
\Big\lbrack\Big( L(p)h,\rho(h^{-1})\xi\left(l_{e^{\epsilon A}}(p)\right)
\Big)\Big\rbrack \nonumber \\ & \hspace{-70pt} \hspace{34pt} - {}
\Big\lbrack\Big(L(p)h, \rho(h^{-1})\xi(p) \Big)\Big\rbrack \Big)
\nonumber \\
& \hspace{-70pt} = \lim_{\epsilon \to 0} \frac{1}{\epsilon}
\Big\lbrack\Big( L(p),\xi\left(l_{e^{\epsilon A}}(p)\right) - \xi(p)
\Big)\Big\rbrack  \nonumber \\
& \hspace{-70pt} = \mathbbm{L}^{(L)}_{\pi_* A^\flat} [(L(p),\xi(p))].
\nonumber
\end{align}
On the other hand we find for the transformation of
$\mathbbm{L}^{(L)}_{\pi_* A^\flat}$ under gauge transformations
\begin{align}
\mathbbm{L}^{(Lh)}_{\pi_* A^\flat} s(p) & = \lim_{\epsilon \to 0}
\frac{1}{\epsilon} \left( \Big\lbrack\Big( L^{(Lh)}_{e^{-\epsilon A}}
\left(L(l_{e^{\epsilon A}}(p))\right),\xi\left(l_{e^{\epsilon
A}}(p)\right) \Big)\Big\rbrack -
\Big\lbrack\Big(L(p) , \xi(p) \Big)\Big\rbrack \right) \nonumber \\
&  = \lim_{\epsilon \to 0} \frac{1}{\epsilon} \Big( \Big\lbrack\Big( L(p)
h(p),\rho(h(l_{e^{\epsilon A}}(p))^{-1})\xi\left(l_{e^{\epsilon
A}}(p)\right) - \rho(h(p)^{-1})\xi(p)
\Big)\Big\rbrack  \nonumber \\
&  = \tilde{R}^{(L)}_{h(p)} \lim_{\epsilon \to 0} \frac{1}{\epsilon} \Big(
\Big\lbrack\Big( L(p) ,\rho(h(l_{e^{\epsilon
A}}(p))^{-1})\xi\left(l_{e^{\epsilon A}}(p)\right) - \rho(h(p)^{-1})\xi(p)
\Big)\Big\rbrack  \nonumber \\
&  = \tilde{R}^{(L)}_{h(p)} \,\, \mathbbm{L}^{(L)}_{\pi_* A^\flat} \left(
\tilde{R}^{(L)}_{h(p)^{-1}} s(p)\right) \nonumber,
\end{align}
i.e.\
\begin{equation}
\mathbbm{L}^{(Lh)}_{\pi_* A^\flat} = \tilde{R}^{(L)}_{h(p)} \,\,
\mathbbm{L}^{(L)}_{\pi_* A^\flat} \tilde{R}^{(L)}_{h(p)^{-1}}.
\end{equation}

From \refe{eqn:hcovpi*} we easily find for the algebra of two
$H$-covariant derivatives
\begin{equation}
[\mathbbm{L}^{(L)}_{\pi_* A^\flat},\mathbbm{L}^{(L)}_{\pi_* B^\flat}] =
\mathbbm{L}^{(L)}_{[\pi_* A^\flat,\pi_* B^\flat]}.
\end{equation}

Now, from \refe{eqn:hcovpi*} we immediately find that the $H$-covariant
Lie derivative of the frame $E_A^{(L)}$ gives zero
\begin{equation}
\mathbbm{L}^{(L)}_{\pi_* A^\flat} E_A^{(L)}  = \Big\lbrack \Big(L(p)\,,\,
\left\{\pi_* A^\flat [\delta_A{}^B]\right\}\,\Big) \Big\rbrack = 0.
\end{equation}
The $H$-covariant Lie derivative of the spin connection $\omega^{(L)}$ is
given by
\begin{align}
\mathbbm{L}^{(L)}_{\pi_* A^\flat} \omega^{(L)} & = \Big\lbrack \Big(L(p),
\left\{\pi_* A^\flat
[\omega^{(L)}_A]\right\} \Big) \Big\rbrack \nonumber \\
& = \Big\lbrack \Big(L(p), \left\{E_A^{(L)}[W_L^{(L)}(A)] -
[W_L^{(L)}(A),\omega_A^{(L)}] + W_L^{(L)I}(A) f_{IA}{}^B \omega^{(L)}_B
\right\}\Big) \Big\rbrack,
\end{align}
where we have used the infinitesimal composition rule for the left and
right $H$-compensators, \refe{eqn:leftrightcomprule1}.

\subsection{Comparison to the literature}
$H$-covariant Lie derivatives have been mentioned in various places in the
literature, see for example \cite{coset,lightblue}. In this section we
shall show how our definition relates to those given in the literature. In
order to do this we shall first derive a relation between the ordinary Lie
derivative and the $H$-covariant Lie derivative on a general section of an
associated bundle.

Recall the expression for the Lie derivative given in \refe{eqn:liecomp},
\begin{displaymath}
\mathcal{L}_{\pi_* A^\flat} s(p)  = \Big\lbrack \Big(\,L(p)\,,\, \pi_*
A^\flat [\xi(p)] + \rho\big(W^{(L)}_L(A)\big) \xi(p)\,\Big) \Big\rbrack.
\end{displaymath}
Using the definition of the action of the algebra of $H$ on a local
section, see \refe{eqn:deftildeRH}, and the expression for the
$H$-covariant Lie derivative given in \refe{eqn:hcovpi*} we find from this
\begin{equation}
\label{eqn:rel_lieversliecomp} \mathcal{L}_{\pi_* A^\flat} s(p)  = \big(
\mathbbm{L}^{(L)}_{\pi_* A^\flat} + \tilde{R}^{(L)}_{W^{(L)}_L(A)}\big)
s(p).
\end{equation}
From this we see that the Lie derivative consists of two parts, namely the
$H$-covariant Lie derivative and a gauge transformation. Let us understand
this in more detail.

\begin{figure}[t]
\begin{center}
\epsfig{file=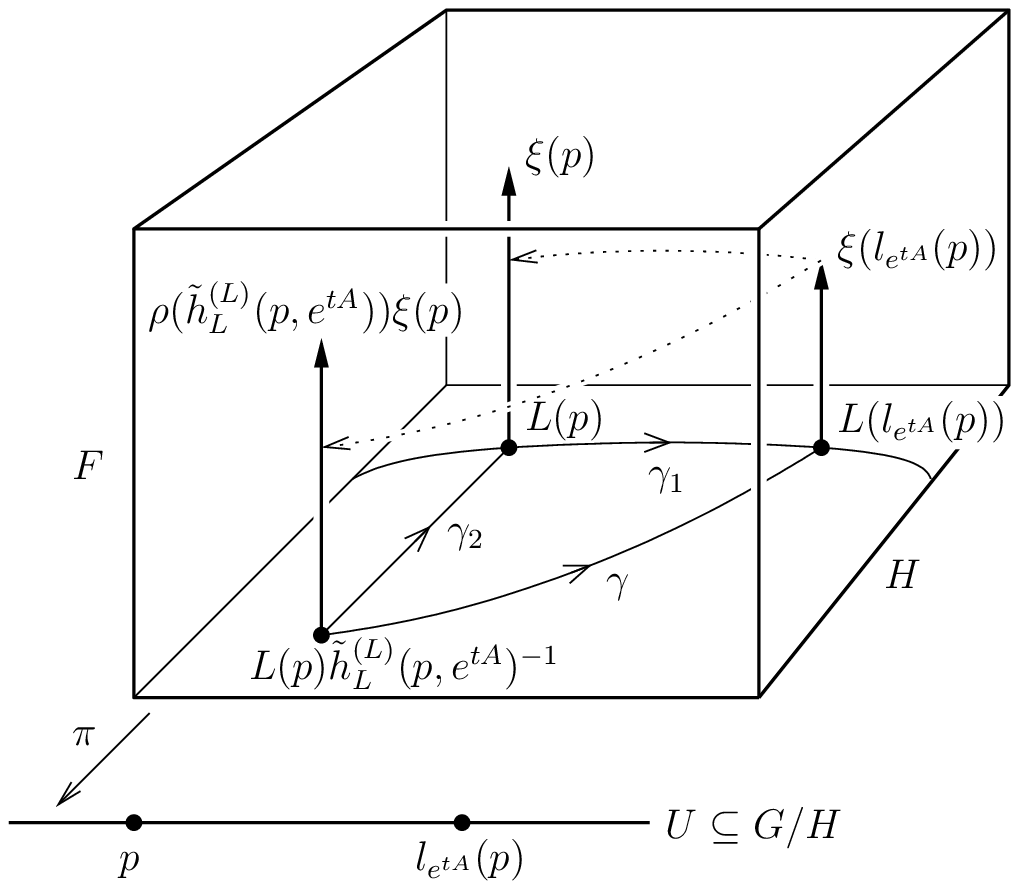}
\end{center}
\caption[]{Illustration of the dragging operation associated with the Lie
derivative and the $H$-covariant Lie derivative, respectively.}
\label{fig:drag}
\end{figure}

First consider the $H$-covariant Lie derivative on a local section $s(p) =
[(L(p), \xi(p))]$.  This measures the difference between
$\xi(l_{e^{tA}}(p))$ at $L(l_{e^{tA}}(p))$ and $\xi(p)$ at $L(p)$ by
dragging $\xi(l_{e^{tA}}(p))$ back constantly along the curve
$\gamma_1(s_1) = L_{e^{s_1A}}^{(L)}(L(p))$, see Figure \ref{fig:drag}.

In contrast, the ordinary Lie derivative measures the difference between
$\xi(l_{e^{tA}}(p))$ at $L(l_{e^{tA}}(p))$ and $\xi(p)$ at $L(p)$ by
dragging $\xi(l_{e^{tA}}(p))$ back constantly along the curve
$\gamma(s)=L_{e^{sA}}(L(p)\tilde{h}_L^{(L)}(p,e^{tA})^{-1})$.
Alternatively this can be understood as dragging $\xi(l_{e^{tA}}(p))$ back
constantly first along the local section $L(p)$, namely along the curve
$\gamma_1(s_1)$, and then along the curve
$\gamma_2(s_2)=R_{\tilde{h}_L^{(L)}(p,e^{s_2A})}(L(p)\tilde{h}_L^{(L)}(p,e^{tA})^{-1})$,
see Figure \ref{fig:drag}.

As such the Lie derivative splits up into two parts: The first one, the
$H$-covariant Lie derivative, measures the change along the local section
$L(p)$, the second one measures the change in the direction
`perpendicular' to the local section $L(p)$.

Now consider the $H$-covariant Lie derivative. This is given by, see
\refe{eqn:hcovpi*},
\begin{displaymath}
\mathbbm{L}^{(L)}_{\pi_* A^\flat}s(p) = [(L(p), \pi_* A^\flat [\xi(p)])].
\end{displaymath}
Recall that $\xi(p) \in F$ stands for the components of the supertensor
$s(p)$ with respect to the basis of the $E_A^{(L)}$. E.g., for $s(p) = X$
a vector, $\xi(p)$ stands for the components $X^A$ of this vector in the
basis $E_A^{(L)}$, see Section \ref{sec:supertensorbundles}. As such
$\pi_* A^\flat [\xi(p)]$ is equal to the Lie derivative of the
\emph{components} of the supertensor $s(p)$. E.g., in  the case of $s(p) =
X$ a vector, we have $\pi_* A^\flat[ X^A(p)] = \ell_{\pi_* A^\flat}
X^A(p)$, where we wrote the Lie derivative of the \emph{scalar} quantity
$X^A$ as $\ell_{\pi_* A^\flat} X^A(p)$ in order to distinguish this from
$(\mathcal{L}_{\pi_* A^\flat}X)^A$, i.e.\ from the $A$th component of the
Lie derivative of $X$. Note that the difference we make between the Lie
derivative $\ell_{\pi_* A^\flat}$ and the usual Lie derivative
$\mathcal{L}_{\pi_* A^\flat}$ is a purely notational one.

We can thus rewrite the $H$-covariant Lie derivative as
\begin{equation}
\mathbbm{L}^{(L)}_{\pi_* A^\flat}s(p) = [(L(p), \ell_{\pi_* A^\flat}.
\xi(p))]
\end{equation}

Now let us consider the Lie derivative. Using the above notation we can
define
\begin{equation}
\label{eqn:hcovliterature} \mathbbm{l}^{(L)}_{\pi_* A^\flat} \defined
\ell_{\pi_* A^\flat}  + \rho\big(W^{(L)}_L(A)\big),
\end{equation}
where $ \rho\big(W^{(L)}_L(A)\big)$ transforms the tangent space indices
in the way following from \refe{eqn:equivreltensor2}. This allows us to
rewrite the Lie derivative of a local section $s(p)$ as
\begin{equation}
\mathcal{L}_{\pi_* A^\flat} s(p) = [(L(p),\mathbbm{l}^{(L)}_{\pi_*
A^\flat}\xi(p))].
\end{equation}
In the literature it is $\mathbbm{l}^{(L)}_{\pi_* A^\flat}$, defined in
\refe{eqn:hcovliterature}, that is commonly referred to as the
$H$-covariant Lie derivative. One should note the similarity of
\refe{eqn:hcovliterature} to the expression for the Lie derivative in
terms of the $H$-covariant Lie derivative, see
\refe{eqn:rel_lieversliecomp}.

As an example consider the Lie derivative and the $H$-covariant Lie
derivative, respectively, of $s(p) = X$ a vector. We find
\begin{subequations}
\begin{align}
(\mathcal{L}_{\pi_* A^\flat} X)^A & = \mathbbm{l}^{(L)}_{\pi_* A^\flat} (X^A) \\
(\mathbbm{L}^{(L)}_{\pi_* A^\flat} X)^A & = \ell_{\pi_* A^\flat} (X^A),
\end{align}
where
\begin{equation}
\mathbbm{l}^{(L)}_{\pi_* A^\flat} (X^A) = \pi_* A^\flat [X^A] + X^B
W_L^{(L)I}(A)f_{IB}{}^A.
\end{equation}
\end{subequations}

Now consider the covariant derivative. We have, see
\refe{eqn:covdercomp1},
\begin{displaymath}
\nabla_A^{(L)} s(p) = \Big\lbrack \Big(\,L(p)\,,\, E_A^{(L)} [\xi(p)] +
\rho\big(\omega^{(L)}_A\big) \xi(p)\,\Big) \Big\rbrack
\end{displaymath}
and setting
\begin{equation}\label{eqn:Ddef}
\mathcal{D}^{(L)}_A \defined E_A^{(L)} + \rho\big(\omega^{(L)}_A\big)
\end{equation}
we can rewrite the covariant derivative as
\begin{equation}
\nabla_A^{(L)} s(p) = \Big\lbrack \Big(\,L(p)\,,\, \mathcal{D}_A^{(L)}
\xi(p)\,\Big) \Big\rbrack.
\end{equation}
Now, considering the commutator of the covariant derivative with the Lie
derivative we have from \refe{eqn:commliecov0}
\begin{displaymath}
[\mathcal{L}_{\pi_* A^\flat}, \nabla_{E_A}]s(p)  = \Big[\Big(L(p),
[\mathbbm{l}^{(L)}_{\pi_* A^\flat}, \mathcal{D}^{(L)}_A]\xi(p)\Big)\Big] =
0
\end{displaymath}
and hence
\begin{equation}\label{eqn:lDcommutator}
[\mathbbm{l}^{(L)}_{\pi_* A^\flat}, \mathcal{D}^{(L)}_A] = 0.
\end{equation}

\section{Killing supervectors}\label{sec:killing}
We have, in Section \ref{sec:isometries}, touched upon the notion of
Killing supervectors, the supervectors associated to infinitesimal
isometries. In this section we shall analyze in more detail how the
Killing supervectors may be defined, particularly in terms of the
derivations we discussed in Section \ref{sec:derivations}.

\subsection{The generalized Lie derivative}
Let us first introduce what we shall call the \emph{generalized Lie
  derivative}. Just as the Lie derivative and covariant derivative
  were constructed from the action of certain local
  bundle maps so will the generalized Lie derivative. The map we use
  will be a local bundle map of the form
\begin{align}
F_t^{(L)} &: \pi^{-1}(U) \to \pi^{-1}(U) \nonumber \\
 &: L(p)h \mapsto L(f_t(p))\tilde{h}_t(p)^{-1}h.
\end{align}
Here $f_t:U \to U$ is a one parameter family of maps on the base and the
maps $\tilde{h}_t:U \to H$ are a one parameter family of $H$-valued
functions. The map $F_t^{(L)}$ is required to satisfy
$F_0^{(L)}=\mathrm{id}_{\pi^{-1}(U)}$, and so, associated with the map is
a supervector field on $\pi^{-1}(U)$ defined by
\begin{equation}
V|_g[f]=\dd{t}f(F_t^{(L)}(g))\Big|_{t=0}.
\end{equation}
Following from the decomposition of the map $F_t^{(L)}$ as mentioned above
we have a decomposition of the supervector field $V$ as
\begin{equation}\label{eqn:Vdecomp}
V|_{L(p)}=L_\ast \big(X|_p\big) - H^\sharp|_{L(p)}.
\end{equation}
Here $X|_p=\pi_\ast \big(V|_{L(p)}\big)$ is the supervector associated to
the map $f_t$ on the base and $H$ is the algebra element associated to
$\tilde{h}_t(p)$ given by $\tilde{h}_\epsilon (p) = 1 + \epsilon H(p) +
\mathcal{O}(\epsilon^2)$. It can be shown that the supervector field $V$
at points off the local section is given by
\begin{equation}
V|_{L(p)h} = {R_h}_\ast \left( V|_{L(p)} \right).
\end{equation}
As a consequence of this $\pi_\ast V$ is a well defined supervector field
on the base manifold, without the need to specify from which point the
supervector $V$ was taken to originate, analogously to the situation for
$\pi_\ast A^\flat$.

The map $F_t^{(L)}$ can then be extended to a map on an associated bundle
analogously to \refes{eqn:tildeLg}{eqn:tildeRgL}
\begin{equation}
\tilde{F}_t^{(L)}:[(L(p),\xi(p))] \mapsto [(F_t^{(L)}(L(p)),\xi(p))].
\end{equation}
We then define the generalized Lie derivative of a (local) section $s(p)$
of the associated bundle by
\begin{equation}
\mathcal{K}_V s(p) \defined \lim_{\epsilon\to 0} \frac{1}{\epsilon} \left(
\left(\tilde{F}_\epsilon^{(L)}\right)^{-1} s(f_\epsilon(p)) - s(p)
\right).
\end{equation}
Note that here we have indexed the generalized Lie derivative with the
supervector $V$. To see how $\mathcal{K}_V$ depends on the components $X$
and $H$, first note that
\begin{equation}
\big(F_t^{(L)}\big)^{-1}L(f_t(p))=L(p)\tilde{h}_t(p).
\end{equation}
Using this we find from the definition of $\mathcal{K}_V$ that
\begin{align}
\nonumber \mathcal{K}_V s(p) &= \lim_{\epsilon\to 0} \frac{1}{\epsilon}
\left(\left[\Big( L(p)\tilde{h}_\epsilon(p) ,
  \xi(f_\epsilon(p))\Big)\right] - \left[\Big( L(p),
  \xi(p)\Big)\right]\right)  \\
\nonumber &= \lim_{\epsilon\to 0} \frac{1}{\epsilon} \Big( \Big[\Big( L(p)
, \xi(f_\epsilon(p)) - \xi(p)\Big)\Big]
\\
\nonumber &\hspace{2cm}+ \Big[\Big( L(p) , (\rho(\tilde{h}_\epsilon(p))-1)
  \xi(f_\epsilon(p))\Big)\Big]\Big) \\
\label{eqn:KXsplit} &= \big(\mathbbm{L}^{(L)}_X + \tilde{R}^{(L)}_H \big)
s(p).
\end{align}
In the last line we have introduced a generalization of the $H$-covariant
Lie derivative, see \refe{eqn:hcovdef}, to act in the direction of an
arbitrary supervector field $X$ rather than just in the direction of a
Killing supervector $\pi_\ast A^\flat$
\begin{equation}
\mathbbm{L}^{(L)}_X s(p) \defined \big[\big(L(p),X[\xi(p)] \big)\big].
\end{equation}
Also we have used the map $\tilde{R}_H^{(L)}$ defined in
\refe{eqn:deftildeRH} for the algebra element $H$ defined by
$\tilde{h}_\epsilon (p) = 1 + \epsilon H(p) + \mathcal{O}(\epsilon^2)$.
\refe{eqn:KXsplit} should be compared to the similar result for the Lie
derivative, \refe{eqn:rel_lieversliecomp}. Note that $\mathcal{K}_V$ may
or may not depend on the local section, we will not explicitly indicate
this dependence.

The generalized Lie derivatives can be shown to form an algebra
\begin{equation}
[\mathcal{K}_U,\mathcal{K}_V]=\mathcal{K}_{[U,V]}.
\end{equation}
For the supervector $U$ we will write the decomposition,
\refe{eqn:Vdecomp}, as $U=L_\ast X_U - H_U^\sharp$, and similarly for the
supervectors $V$ and $[U,V]$. Then, the algebra of the generalized Lie
derivatives decomposes as
\begin{subequations}
\begin{gather}
\label{eqn:genderivalgcomp1} X_{[U,V]}=[X_U,X_V] \\
\label{eqn:genderivalgcomp2} H_{[U,V]}=X_U[H_V] - X_V[H_U] + [H_U,H_V].
\end{gather}
\end{subequations}
These equations can be seen to encompass the various properties of the
left and right $H$-compensators discussed in Section
\ref{sec:propertiesHcompensators}. For instance, if we take the
supervectors $U=A^\flat$ and $V=B^\flat$ then $[U,V]=[B,A]^\flat$. The
decomposition gives $X_U=\pi_\ast A^\flat$ and $H_U=W_L^{(L)}(A)$, and
similarly for $V$ and $[U,V]$. Using these values we see that
\refec{eqn:genderivalgcomp1}{eqn:genderivalgcomp2} give precisely
\refec{eqn:algebrapiflats}{eqn:intcondW1}. Following a similar approach
with more elaborate choices of the supervectors $U$ and $V$ we may
likewise obtain \refec{eqn:rginf}{eqn:hrinf} and
\refec{eqn:leftrightcomprulemaps1}{eqn:leftrightcomprule1}.

We see from \refe{eqn:KXsplit} that the action of $\mathcal{K}_V$ can be
thought of as comprised of two parts. The first part,
$\mathbbm{L}^{(L)}_X$, can be thought of as arising from an infinitesimal
general coordinate transformation in the direction of the supervector $X$.
The second part, $\tilde{R}^{(L)}_H$, can be thought of as arising from an
infinitesimal local gauge transformation. Thus $\mathcal{K}_V$ can be
thought of as giving the general infinitesimal transformation of a
section. This is also apparent from the form of the map $F_t^{(L)}$ that
was used to define $\mathcal{K}_V$.

\subsection{Definition of Killing supervectors}
We seek a definition of Killing supervectors which is consistent with our
notion of infinitesimal isometries as discussed in Section
\ref{sec:isometries}. Since $\mathcal{K}_V$ represents a general
infinitesimal transformation we would like to find a condition on
$\mathcal{K}_V$ such that it represents an infinitesimal isometry. Once we
have this, then the supervector field $X$ in the decomposition
\refe{eqn:Vdecomp} will be the desired Killing supervector.

Recall from Section \ref{sec:isometries} that we defined isometries to be
the group of transformations which leave the supergeometry, \ie the frame
and connection, invariant up to a single gauge transformation. Now the
covariant derivative is a quantity which contains both the frame and the
connection, thus we would suspect that there is some way of utilizing the
covariant derivative to give a condition for $\mathcal{K}_V$ to be an
infinitesimal isometry.

Now, recall that the Lie derivative in the direction $\pi_\ast A^\flat$
commutes with the covariant derivative, \refe{eqn:commliecov0}. With this
in mind let us consider the commutator of the covariant derivative with
the generalized Lie derivative. We have
\begin{align}
\nonumber [\mathcal{K}_V , \nabla^{(L)}_A] s(p) &= \mathcal{K}_V
\big[\big(
  L(p), (E^{(L)}_A + \omega^{(L)}_A) \xi(p) \big)\big] \\
\nonumber &\hspace{1cm}- \nabla^{(L)}_A \big[\big( L(p), (X + H)
  \xi(p) \big)\big]  \\
\nonumber &= \Big[\Big( L(p), \Big([X,E^{(L)}_A] - V^I f_{IA}{}^B
  E^{(L)}_B + X[\omega^{(L)}_A] - E^{(L)}_A[H]  \\
\label{eqn:KXdelcommutator} &\hspace{1cm}  + [H,\omega^{(L)}_A]
  - V^If_{IA}{}^B\omega^{(L)}_B\Big) \xi(p) \Big)\Big],
\end{align}
where we have taken the algebra valued quantities $\omega_A^{(L)}$ and
$H=V^IH_I$ to be acting in the appropriate representation. Thus supposing
we impose the condition
\begin{equation}\label{eqn:killingcommutator}
 [\mathcal{K}_V , \nabla^{(L)}_A] = 0
\end{equation}
then we see from \refe{eqn:KXdelcommutator} that this implies
\begin{subequations}
\begin{align}
\label{eqn:LXEA} [X,E^{(L)}_A] &= V^I f_{IA}{}^B E^{(L)}_B, \\
\label{eqn:LXomegaA} X[\omega^{(L)}_A] - V^If_{IA}{}^B\omega^{(L)}_B &=
E^{(L)}_A[H] - [H,\omega^{(L)}_A].
\end{align}
\end{subequations}
Now, we note that\footnote{Here we shall use the definition of the Lie
  derivative $\mathcal{L}_X$ (and $\ell_X$) in an arbitrary direction
  $X$ as opposed to $\pi_\ast A^\flat$, this
  is defined in the usual way when acting on vectors, scalars and forms.} $\mathcal{L}_X E^{(L)}_A = [X,E^{(L)}_A]$, thus the
first equation gives us an expression for the Lie derivative of the frame.
Also we have
\begin{align}
\nonumber X[\omega^{(L)}_A] &= \ell_X \big(\omega^{(L)}(E^{(L)}_A)\big) \\
\nonumber &= \big(\mathcal{L}_X\omega^{(L)}\big)(E^{(L)}_A) +
\omega^{(L)} \big( \mathcal{L}_X E^{(L)}_A \big) \\
\label{eqn:XomegaA} &= \big( \mathcal{L}_X\omega^{(L)} \big) (E^{(L)}_A) +
V^I f_{IA}{}^B \omega^{(L)}_B,
\end{align}
where in the last line we have used the expression for $\mathcal{L}_X
E^{(L)}_A$ given by \refe{eqn:LXEA}. Using \refe{eqn:XomegaA} we see that
\refe{eqn:LXomegaA} gives us an expression for the $A$th component of the
Lie derivative of $\omega^{(L)}$. Thus in total we may rewrite
\refec{eqn:LXEA}{eqn:LXomegaA} as
\begin{subequations}
\begin{align}
\mathcal{L}_X E^{(L)}_A &=  V^I f_{IA}{}^B E^{(L)}_B \\
\mathcal{L}_X \omega^{(L)} &= \extd H - [H,\omega^{(L)}].
\end{align}
\end{subequations}
These equations are simply the first order contributions to
\begin{subequations}
\begin{align}
{f_\epsilon}_\ast \left(E^{(L)}_A\big|_p \right) &=
\Lambda_A{}^B(\tilde{h}_\epsilon) E_B^{(L)}\big|_{f_\epsilon(p)} \\
{f_\epsilon}^\ast \left(\omega^{(L)}\big|_{f_\epsilon(p)}\right) &=
\tilde{h}_\epsilon^{-1} \omega^{(L)}\big|_p \tilde{h}_\epsilon +
\tilde{h}_\epsilon^{-1} \extd \tilde{h}_\epsilon.
\end{align}
\end{subequations}
If we compare this result to \refec{eqn:isome}{eqn:isomomega} we see that
infinitesimally the transformation $f_\epsilon$ represents an isometry, as
under its action the frame and connection both transform by the gauge
transformation given by $\tilde{h}_\epsilon$.

Based on these considerations we take \refe{eqn:killingcommutator} as
defining the Killing supervectors: The supervector $X=\pi_\ast V$
associated with a transformation $\mathcal{K}_V$ is a Killing supervector
if the transformation  $\mathcal{K}_V$ commutes with the covariant
derivative $\nabla^{(L)}_A$. The advantage of this definition of Killing
supervectors is that it is easily generalized to superspaces which are not
written as a coset space. So long as we have a covariant derivative we
just seek the infinitesimal combined coordinate and gauge transformation
which commutes with it. This is the approach taken in \cite{bluebook},
however there the derivations used are viewed as acting on the components
of a section rather than on the section itself.

\subsection{Killing supervectors for the super coset space}
We know from Section \ref{sec:isometries} that the supervector fields
$\pi_\ast A^\flat$ are Killing supervectors of the coset space $G/H$. We
would like to see how we may obtain these Killing supervectors from
\refe{eqn:killingcommutator}. Let us first start by rewriting
\refe{eqn:killingcommutator} as, cf.\ \refe{eqn:commutatorgradlie},
\begin{equation}\label{eqn:altkillingcommutator}
[\mathcal{K}_V,\nabla_{\pi_\ast B^\sharp}] =
\nabla_{(\mathcal{K}_V\pi_\ast B^\sharp)}, \qquad \forall B
\in\mathfrak{k}
\end{equation}
which is easily checked noting that $B=Y^A K_A$ for some constants $Y^A$.
Note we again here use the shorthand of dropping the dependence on $L(p)$
in the vector $\pi_\ast B^\sharp$. The supervector $V$ will be decomposed
into components $X$ and $H$ as in \refe{eqn:Vdecomp}. Then as the $Y^A$
are constants we have that $\mathbbm{L}_X^{(L)} \pi_\ast B^\sharp = 0$,
hence
\begin{equation}
\mathcal{K}_V \pi_\ast B^\sharp = \tilde{R}_H^{(L)} \pi_\ast B^\sharp =
\pi_\ast [H,B]^\sharp.
\end{equation}
Then, either by direct calculation or by using
\refec{eqn:genderivalgcomp1}{eqn:genderivalgcomp2}, we see that
\refe{eqn:altkillingcommutator} decomposes into the following two
conditions
\begin{subequations}
\begin{gather}
\label{eqn:kc1} \pi_\ast [H,B]^\sharp = [X,\pi_\ast B^\sharp ] \\
\label{eqn:kc2} \omega^{(L)}(\pi_\ast[H,B]^\sharp) =
X[\omega^{(L)}(\pi_\ast
  B^\sharp)] - \pi_\ast B^\sharp [H] + [H,\omega^{(L)}(\pi_\ast B^\sharp)],
\end{gather}
\end{subequations}
which must be satisfied for all possible choices of the algebra element
$B$. We know from
\refec{eqn:leftrightcomprulemaps1}{eqn:leftrightcomprule1} that a solution
to these conditions is $X=\pi_\ast A^\flat$ and $H=W_L^{(L)}(A)$ for some
algebra element $A$. We shall now show that all solutions to these
equations can be expressed in this form. Stated differently, we will show
that the unique solution to \refe{eqn:killingcommutator} is given by
\begin{equation}
\mathcal{K}_V = \mathcal{L}_{\pi_\ast A^\flat},
\end{equation}
and the Killing supervectors are thus of the form $\pi_\ast A^\flat$.

Let us expand $X=X^AE_A$ and $H=V^IH_I$, then \refec{eqn:kc1}{eqn:kc2} may
be rewritten as
\begin{align}
\nonumber \pi_\ast B^\sharp [X^A] &= X^B [E_B, \pi_\ast B^\sharp]^A -
V^I(\pi_\ast[H_I,B]^\sharp)^A \\
\nonumber \pi_\ast B^\sharp [V^I] &= X^AE_A[\omega^I(\pi_\ast B^\sharp)] +
V^J[H_J,\omega(\pi_\ast B^\sharp)]^I -
V^J\omega(\pi_\ast[H_J,B]^\sharp)^I,
\end{align}
which should be satisfied for all choices of $B\in\mathfrak{k}$. As usual
we drop the dependence on $L(p)$ for the vectors $\pi_\ast B^\sharp$.
Thus, if we group the components into a single object $Z^p=(X^A,V^I)$
these equations take the simple form
\begin{equation}\label{eqn:compactkc}
\pi_\ast B^\sharp [Z^p] + Z^qM_q{}^p(B) = 0, \qquad \forall
B\in\mathfrak{k}
\end{equation}
where the objects $M_q{}^p(B)$ depend linearly on the choice of $B$, they
are also functions of the frame and connection. This equation is linear
and homogeneous in $Z$, thus solutions may be combined linearly to give
other solutions. Clearly also $Z^p=0$ is a solution for all choices of
$B$. In fact we already have a whole family of solutions parameterized by
the algebra elements $A\in\mathfrak{g}$, we will denote these solutions by
$Z^p(A)$ whose components are $X(A)=\pi_\ast A^\flat$ and
$H(A)=W_L^{(L)}(A)$. We will now show that all solutions are of this type.

Suppose we have some solution $Z^p$ to \refe{eqn:compactkc}. Let us focus
on a particular point, $p$, in the base manifold and consider a curve
$\gamma(t)$ joining the point $p=\gamma(0)$ to some nearby point
$q=\gamma(1)$. At each point $\gamma(t)$ along the curve the tangent to
the curve can be expressed as $\pi_\ast B(t)^\sharp$ for some choice of
algebra element $B(t)$, this can be viewed as simply the expansion of the
tangent vector in the basis $E_A$. Now, our solution $Z^p$ to
\refe{eqn:compactkc} is a solution for all choices of $B$ at all points,
and thus in particular the choice $B(t)$ at the point $\gamma(t)$.
Abbreviating $Z^p|_{\gamma(t)}=Z^p(t)$ and $M_q{}^p(B(t))=M_q{}^p(t)$
\refe{eqn:compactkc} becomes, for this specific choice,
\begin{equation}\label{eqn:diffkc}
\dd{t} Z^p(t) + Z^q(t)M_q{}^p(t) = 0
\end{equation}
which should then be satisfied for all $t$ along the curve.

Now, \refe{eqn:diffkc} is a linear first order differential equation.
Thus, given the initial conditions, $Z^p(0)$, the solution $Z^p(t)$ is
uniquely determined for $t>0$. In particular $Z^p(1)$ is uniquely
determined. In fact, since the curve $\gamma$ and the point $q$ are
arbitrary we only need to specify $Z^p$ at one single point in the base
manifold from which $Z^p$ will be uniquely determined for all other
points. Thus, to show that all solutions to \refe{eqn:compactkc} can be
expressed in the form $Z^p(A)$ for some algebra element $A$ it is
sufficient to show that at a single point $p$ an arbitrary $Z^p$ can be
expressed as $Z^p(A)$ for some $A$.

So let us consider a solution $Z^p$ at the point $p$ composed of some
supervector $X|_p$ and algebra element $H|_p\in\mathfrak{h}$. First note
that it is always possible to choose an algebra element $A\in\mathfrak{g}$
such that
\begin{equation}\label{eqn:choiceofAvector}
\pi_\ast \left(A^\flat|_{L(p)}\right) = X|_p.
\end{equation}
For instance it is straightforward to show that the choice
$A=X|_p[L]L(p)^{-1}$ satisfies this equation. However, this choice of $A$
is not unique, there is a freedom in the choice of $A$ under which
$\pi_\ast A^\flat$ at the point $p$ remains unchanged. This can be seen in
the following way. Let $\tilde{H}$ be an arbitrary element of
$\mathfrak{h}$, then
\begin{align}
\nonumber L(l_{e^{tA}}(p))\tilde{h}_L^{(L)}(p,e^{tA}) &= e^{tA} L(p) \\
\nonumber &=  e^{tA}e^{tL\tilde{H}L^{-1}} L(p) e^{-t\tilde{H}} \\
\nonumber &= e^{t(A+L\tilde{H}L^{-1})} L(p) e^{-t\tilde{H}} + \mathcal{O}(t^2) \\
\nonumber &= L(l_{e^{t(A+L\tilde{H}L^{-1})}}(p))
\tilde{h}_L^{(L)}\big(p,e^{t(A+L\tilde{H}L^{-1})}\big) e^{-t\tilde{H}} +
\mathcal{O}(t^2).
\end{align}
From which we immediately deduce
\begin{subequations}
\begin{gather}
\pi_\ast \big(A^\flat \big|_{L(p)} \big) = \pi_\ast \big(
(A+L(p)\tilde{H}L(p)^{-1})^\flat\big|_{L(p)} \big) \\
W_L^{(L)}\big(p, A+L(p)\tilde{H}L(p)^{-1}\big) = W_L^{(L)}(p,A) +
\tilde{H}.
\end{gather}
\end{subequations}
Therefore, after choosing the algebra element $A$ so that at $p$ we have
$\pi_\ast A^\flat=X$ we still have the freedom to change $A \to
A+L(p)\tilde{H}L(p)^{-1}$, and under such a transformation the
$H$-compensator $W_L^{(L)}(A)$ at $p$ can be changed arbitrarily. Thus as
well as choosing $A$ so that \refe{eqn:choiceofAvector} is satisfied we
may simultaneously choose $A$ such that
\begin{equation}
W_L^{(L)}(p,A) = H|_p.
\end{equation}
This concludes the proof.

\section{Flat superspace}
\label{sec:example} In this section we will apply the concepts discussed
in the previous sections to the well known case of flat superspace. We
start with the super Poincar\'e group $G=S\Pi$ and its subgroup of Lorentz
transformations $H$. Flat superspace is defined to be the coset space
$S\Pi/H$. The non-zero commutators of the super Poincar\'e algebra,
$\mathfrak{g}=\mathfrak{s\pi}$, are
\begin{subequations}
\begin{align}
[M_{ab},M_{cd}] & = \eta_{ad}M_{bc} +
\eta_{bc}M_{ad} - \eta_{ac}M_{bd} - \eta_{bd}M_{ac} \label{eqn:MMcommutator} \\
[M_{ab},P_c] & = \eta_{cb}P_a -\eta_{ca}P_b\label{eqn:MPcommutator}\\
[M_{ab},Q_\alpha] & = - (\sigma_{ab})_{\alpha}{}^\beta Q_\beta \label{eqn:MQcommutator}\\
[Q_\alpha , Q_\beta] & = 2k(\gamma^a C^{-1})_{\alpha\beta}P_a
\label{eqn:QQcommutator}.
\end{align}
\end{subequations}
Here the $\gamma^a$ are the Dirac gamma matrices for the flat metric
$\eta_{ab}$, $\sigma_{ab}=\frac{1}{4}[\gamma_a,\gamma_b]$, $C$ is the
charge conjugation matrix, and $k$ is a phase factor. The indices may be
grouped as $A=(a,\alpha)$ and the antisymmetric pair $I=[ab]$. We will
distinguish between tangent space indices $A=(a,\alpha)$ and coordinate
indices $M=(m,\mu)$, however both sets range over the same values.
Comparing \refes{eqn:MMcommutator}{eqn:QQcommutator} to
\refes{eqn:algebra1}{eqn:algebra3} we have that the even generators
$M_{ab}$ generate the Lorentz subgroup and play the role of the $H_I$
whereas the even generators $P_a$ generating translations and the odd
generators $Q_\alpha$ generating supersymmetry transformations play the
role of the $K_A$. We see that, further to the reductive property,
$\mathfrak{k}$ forms a nilpotent subalgebra of $\mathfrak{g}$. We have
\begin{subequations}
\begin{gather}
\label{eqn:kkk} [\mathfrak{k},\mathfrak{k}] \subseteq \mathfrak{k} \\
\label{eqn:kkk0} [\mathfrak{k},[\mathfrak{k},\mathfrak{k}]]=0.
\end{gather}
\end{subequations}

In the expansion of a general element of the super Poincar\'e algebra,
$X=X^AK_A+X^IH_I$, it is possible to restrict the $X^I$ to be
\emph{ordinary} real numbers. This is a result of the reductive property
of the super Poincar\'e algebra and the following: The $H_I$ are all even,
the structure constants $f_{IJ}{}^K$ are ordinary real numbers (i.e.\ $H$
is conventional \cite{bluebook,dewitt}) and the coset space is flat, i.e.\
the structure constants $f_{AB}{}^I$ vanish, cf.\ \refe{eqn:TRcomponents}.
While such a choice for the super Poincar\'e algebra is not necessary it
does allow one to deal simply with the ordinary Lorentz group.

As a consequence of the reductive property any element $g\in S\Pi$ can be
written as $g=e^{x^MK_M}e^{y^IH_I}$ for some $x^M$ and $y^I$
\cite{bluebook}. Further, using \refe{eqn:kkk}, one can show that $x^M$
and $y^I$ are uniquely determined by $g$. It follows that the principal
bundle is trivial, admitting a global section
\begin{align}
\nonumber L&:S\Pi/H \to S\Pi \\
&:p \mapsto e^{x^M(p)K_M}
\end{align}
where $x^M(p)$ are uniquely determined in terms of $p$ and thus can be
used as the coordinates of the point $p$. This section is horizontal and
can be thought of as a special choice of gauge.

Let us calculate the action of an isometry on these coordinates. Consider
left multiplication by $g=e^A$. For $A=Y^AK_A\in\mathfrak{k}$ we can use
the Baker-Campbell-Hausdorff formula, which combined with \refe{eqn:kkk0}
gives
\begin{displaymath}
e^{Y^AK_A}e^{x^MK_M}=e^{Y^AK_A+x^MK_M+\frac{1}{2}[Y^AK_A,x^MK_M]}.
\end{displaymath}
From \refe{eqn:kkk} we see that under this transformation we do not leave
the section $L$. We thus find
\begin{subequations}
\begin{align}
x^M(l_{e^{Y^AK_A}}(p)) &= x^M(p) + Y^A
\Big(\delta_A{}^M-\frac{1}{2}x^N(p)f_{NA}{}^M\Big) \\
\tilde{h}_L^{(L)}\big(p,e^{Y^AK_A}\big) &= 0 .
\end{align}
\end{subequations}
For the algebra element $A=Y^IH_I\in\mathfrak{h}$ we have
\begin{displaymath}
e^{Y^IH_I}e^{x^MK_M}= e^{x^M\Lambda_M{}^N(e^{-Y^IH_I})K_N}e^{Y^IH_I},
\end{displaymath}
where $\Lambda_M{}^N(e^{-Y^IH_I})$ is the coadjoint representation of
$e^{-Y^IH_I}\in H$, see \refe{eqn:coadjoint}. This gives
\begin{subequations}
\begin{align}
x^M(l_{e^{Y^IH_I}}(p)) &= x^M(p) \Lambda_M{}^N(e^{-Y^IH_I}) \\
\tilde{h}_L^{(L)}\big(p,e^{Y^IH_I}\big) &= e^{Y^IH_I}.
\end{align}
\end{subequations}
As in this case the normalizer of the Lorentz subgroup $N(H)$, see Section
\ref{sec:isometries}, is simply $H$ we may determine all Killing
supervectors by considering the infinitesimal versions of these
transformations. We find
\begin{subequations}
\begin{align}
\pi_\ast P_a^\flat &= \partial_a \\
\pi_\ast Q_\alpha^\flat &= \partial_\alpha - kx^\beta(\gamma^aC^{-1})_{\beta\alpha}\partial_a \\
\pi_\ast M_{ab}^\flat &= x_a\partial_b - x_b\partial_a -
x^\alpha(\sigma_{ab})_\alpha{}^\beta\partial_\beta .
\end{align}
\end{subequations}
It is standard to refer to these vectors as the differential operator
representation of the super Poincar\'e algebra. Note, however, that due to
\refe{eqn:algebrapiflats} the algebra will have an extra minus sign
compared to $\mathfrak{s}\pi$.

A similar analysis for right multiplication leads to an expression for the
vectors $\pi_\ast \big(K_A^\sharp|_{L(p)}\big)$, \ie the frame. We find
\begin{subequations}
\begin{align}
E_a^{(L)} &= \partial_a \\
E_\alpha^{(L)} &= \partial_\alpha +
kx^\beta(\gamma^aC^{-1})_{\beta\alpha}\partial_a .
\end{align}
\end{subequations}

The coframe and the connection may be calculated from the pullback of the
Maurer-Cartan form as in \refe{eqn:theequation}. We have
\begin{align*}
\nonumber L^\ast\big(\zeta|_{L(p)}\big) &= L(p)^{-1}\extd L(p) \\
\nonumber &= e^{-x^MK_M}\extd e^{x^MK_M} \\
\nonumber &= \extd x^MK_M + \frac{1}{2}[\extd x^MK_M, x^NK_N] \\
&= \extd x^M (\delta_M{}^A-\frac{1}{2}x^Nf_{NM}{}^A)K_A.
\end{align*}
We may then read off the coframe and connection
\begin{subequations}
\begin{align}
E^a_{(L)} &= \extd x^a - k \,\extd x^m x^n (\gamma^a C^{-1})_{nm}\\
E^\alpha_{(L)} &= \extd x^\alpha \\
\label{eqn:omega0} \omega^{(L)} &= 0.
\end{align}
\end{subequations}
As $\omega^{(L)}=0$ we see from \refe{eqn:Ddef} that
$\mathcal{D}^{(L)}_A=E^{(L)}_A$. This explains how -- as is usually stated
in the literature -- in flat superspace the covariant derivative is
obtained directly in terms of right multiplication on the coset space. In
general we must obviously consider the map $R_g^{(L)}$ of
\refe{eqn:defRgL}.

The covariant derivative $\mathcal{D}_A^{(L)}$ is usually considered as
the operator which (anti)commutes with the differential operator
representation of the $K_A$, \ie the $\pi_\ast K_A^\flat$, particularly
for $A=\alpha$. This can be seen as a special case of
\refe{eqn:lDcommutator} since the $H$-compensator $W_L^{(L)}(K_A)$
vanishes. Note that the covariant derivative will not commute with the
$\pi_\ast M_{ab}^\flat$ as $W_L^{(L)}(M_{ab}) \ne 0$.

Starting with the frame and connection one can construct the Killing
supervectors of flat superspace using the method discussed in Section
\ref{sec:killing}, \ie by imposing that the commutator of the generalized
Lie derivative with the covariant derivative be zero. This procedure, in
the four-dimensional case, is treated in \cite{bluebook}.

The curvature, \refe{eqn:curvaturedef}, of flat superspace is clearly
zero. The only non-zero components of the torsion are given in terms of
the algebra structure constants, see \refe{eqn:TRcomponents}. They are
\begin{equation}
T_{\alpha\beta}{}^a = 2k(\gamma^aC^{-1})_{\alpha\beta}.
\end{equation}

While flat superspace is a very useful example of a super coset space its
geometry is relatively simple. The same techniques we have discussed here
may, however, be applied directly to more complex geometries. A
particularly well studied super coset space is the $AdS_5\times S^5$
superspace which arises as a coset space of the super Lie group
$SU(2,2|4)$, see for example \cite{hep-th/0007099, hep-th/9812087}. A
lower dimensional example of a super coset space is provided by the
supersphere \cite{supersphere, thepaper} which is a coset space of the
super Lie group $\uosp$.

\section{Conclusions}
We have discussed in detail the geometry of super coset spaces with the
focus on how the geometric structures of the coset space $G/H$ are
inherited from $G$.  While the concepts and methods presented in this
paper apply to coset spaces in general, our main aim has been to analyze
the geometry of \emph{super} coset spaces and their isometries. As such
one important aspect of our work was to review and clarify the notion of
Killing supervectors in the context of super coset spaces. Due to the fact
that the notion of supermetric is not physically relevant for the
construction of superspace the standard definition of isometries in terms
of the metric cannot be applied and must be given in terms of the
supergeometry -- the frame and connection. Although the definition of
Killing supervectors we give is derived from the understanding of the
geometry of coset spaces it clearly extends to more general situations.

\subsection*{Acknowledgements}
The authors would like to thank Professor N.\ S.\ Manton for helpful
comments on the manuscript.

\appendix
\section{Appendix}
\subsection{Conventions for super differential forms}
\label{sec:formconventions} We define a super $n$-form $\phi$ on a
supermanifold $M$ at the point $p$ to be a mapping from $n$ copies of the
tangent supervector space to $\R_\infty$, the real supernumbers, i.e.~
\begin{equation}\label{eqn:formdef}
\phi:T_pM \times \ldots \times T_pM \to \R_\infty.
\end{equation}
The target space can be generalized to any vector space with a
``multiplication'', e.g.~a super Lie algebra. While in this paper we write
the supervectors on which a form acts on the right, its properties are
more conveniently represented with the vectors on the left, we define
\begin{equation}
\phi(X_1,\ldots,X_n) \defined (X_n,\ldots,X_1)\cdot\phi.
\end{equation}
We then require the following properties to be satisfied
\begin{subequations}
\begin{align}
(X+Y,Z,\ldots)\cdot\phi &= (X,Z,\ldots)\cdot\phi + (Y,Z,\ldots)\cdot\phi
\\
(\lambda X,Y,\ldots)\cdot\phi &= \lambda (X,Y,\ldots)\cdot\phi \\
(\ldots,X,Y,\ldots)\cdot\phi &= -(-1)^{XY}(\ldots,Y,X,\ldots)\cdot\phi,
\end{align}
\end{subequations}
where in the last equation the supervectors $X$ and $Y$ must be pure
(i.e.~even or odd). From these relations we may further deduce
\begin{subequations}
\begin{align}
(\ldots,X+Y,Z,\ldots)\cdot\phi &= (\ldots,X,Z,\ldots)\cdot\phi +
(\ldots,Y,Z,\ldots)\cdot\phi \\
(\ldots,X\lambda,Y,\ldots)\cdot\phi &= (\ldots,X, \lambda
Y,\ldots)\cdot\phi.
\end{align}
\end{subequations}

Given a $p$-form $\phi$ and a $q$-form $\psi$ we define the exterior
(wedge) product $\phi\wedge\psi$ by its action on a set of $p+q$ pure
supervectors as
\begin{align}
\nonumber (X_{p+q},\ldots,X_1)\cdot\phi\wedge\psi &\defined \frac{1}{p!q!}
\sum_\sigma
(-1)^{\mathrm{sgn}\,\sigma} (-1)^{\nu_\sigma(X_{p+q},\ldots,X_1)} \\
\nonumber & (-1)^{(X_{\sigma (p+q)}+\ldots+X_{\sigma (p+1)})(X_{\sigma
(p)}+\ldots+X_{\sigma (1)}+\phi)} \\
& (X_{\sigma (p)},\ldots,X_{\sigma (1)})\cdot\phi \, (X_{\sigma
(p+q)},\ldots,X_{\sigma (p+1)})\cdot\psi. \label{eqn:wedgedef}
\end{align}
Here $\sigma$ is a permutation on $p+q$ elements. The quantity
$\nu_\sigma$ is defined as
\begin{equation}
a_{\sigma(1)} \ldots a_{\sigma(n)} =
(-1)^{\nu_\sigma(a_1,\ldots,a_n)}a_1\ldots a_n,
\end{equation}
where the $a_i$, $i=1,\ldots,n$, are pure supernumbers. It is possible to
show that the definition \refe{eqn:wedgedef} does indeed define a
$(p+q)$-form. Further, if $\phi$ and $\psi$ are pure forms we have
\begin{equation}
\phi\wedge\psi = (-1)^{\phi\psi+pq}\psi\wedge\phi,
\end{equation}
where the $\phi$ and $\psi$ occurring in the exponent denote the parities
of $\phi$ and $\psi$, respectively.

In the case when the target space in \refe{eqn:formdef} is generalized to
a super Lie algebra we have an \emph{algebra valued form}. The product
used in the definition \refe{eqn:wedgedef} must be replaced by the Lie
algebra bracket, to indicate this we denote the wedge product instead by
$[\phi,\psi]$. The symmetry of this wedge product has an additional minus
sign
\begin{equation}
[\phi,\psi]= -(-1)^{\phi\psi+pq}[\psi,\phi].
\end{equation}
If we work in a matrix representation of the algebra we could instead use
matrix multiplication as the product and define $\phi\wedge\psi$. While
this latter wedge product does not in general result in an algebra valued
form we do however have the relation
\begin{equation}
[\phi,\psi]=\phi\wedge\psi - (-1)^{\phi\psi+pq}\psi\wedge\phi.
\end{equation}
In particular $\phi\wedge\phi$ is algebra valued.

The exterior derivative is defined initially on 0-forms (functions) by
\begin{subequations}
\begin{equation}
\extd f (X) = X[f],
\end{equation}
for an arbitrary supervector $X$. This definition is then extended to
$n$-forms by requiring the following properties to hold
\begin{align}
\extd (\phi + \chi) &= \extd\phi + \extd\chi \\
\extd (\lambda\phi) &= \lambda \extd\phi \\
\extd (\phi\wedge\psi) &= \extd\phi \wedge \psi + (-1)^p \phi \wedge
\extd\psi \\
\extd^2 &= 0.
\end{align}
\end{subequations}
Here $\phi$ and $\chi$ are $p$-forms, $\psi$ is a $q$-form, and $\lambda$
is an arbitrary supernumber.

Let us suppose we have a basis of one-forms $E^A_{(L)}$,
$A=1,\ldots,\mathrm{dim}\,M$. Our conventions for expanding the form
$\phi$ in this basis are, cf.\ \refe{eqn:covariantexp}
\begin{equation}\label{eqn:formexpansion}
\phi \defined \frac{1}{p!} (-1)^{\Delta_p(A,A)} E^{A_1}_{(L)} \wedge
\ldots \wedge E^{A_p}_{(L)} \phi_{A_1 \cdots A_p}.
\end{equation}
Here we have used the parity function, defined as
\begin{equation}\label{eqn:parityfunction}
\Delta_p(A,B) =  \sum_{\stackrel{\scriptstyle t,u}{\scriptstyle t<u}}^p
A_t B_u,
\end{equation}
where $A_t$ and $B_u$ represent the parities of the indices. This function
could in fact take as arguments any set of objects with parity, for
example see \refe{eqn:formvectorcomponents} below. Note that this function
is clearly linear in both arguments. Further, $\Delta_p(A,A)$ is just the
sum over all non-equal pairs of indices and is invariant under any index
permutation. Using this it is possible to show that the components of
$\phi$ are given by
\begin{equation}
\phi_{A_1 \cdots A_p} = (-1)^{\Delta_p(A,A)} (E_{A_p}^{(L)}, \ldots,
E_{A_1}^{(L)})\cdot \phi.
\end{equation}
We can then also show that
\begin{equation}\label{eqn:formvectorcomponents}
(X_p,\ldots,X_1)\cdot\phi = (-1)^{\Delta_p(X,A)}X_p^{A_p}\cdots
X_1^{A_1}\phi_{A_1\cdots A_p}.
\end{equation}

\subsection{Proof of \refe{eqn:Xmakehorizontal} and
  \refe{eqn:Xchangesection}}\label{sec:appendix1}
Consider two curves $\gamma_1,\gamma_2:[0,1]\to G$ in the group which both
project down to the same curve $\gamma:[0,1]\to G/H$ in the base,
i.e.~$\pi\circ\gamma_i=\gamma$, for $i=1,2$. Clearly then we have
\begin{equation}
\gamma_2(t)=\gamma_1(t)h(t)
\end{equation}
for some function $h(t)\in H$. Let $X_1$, $X_2$ denote the tangent vectors
to the curves in the group, and $X$ the tangent vector to the curve in the
base. Working in a matrix representation we thus have
\begin{align}
\nonumber X_2\big|_{\gamma_2(0)} &= \dd{t}\gamma_2(t)\Big|_{t=0} \\
\nonumber &= \dd{t}(\gamma_1(t)h(t))\Big|_{t=0} \\
\nonumber &= \dd{t}\gamma_1(t)\Big|_{t=0} h(0) +  \gamma_1(0)
\dd{t}h(t)\Big|_{t=0} \\
\label{eqn:bothterms} &= \dd{t}(\gamma_1(t)h(0))\Big|_{t=0} + \gamma_2(0)
h(0)^{-1}\dd{t}h(t)\Big|_{t=0}.
\end{align}
Now consider
\begin{align}
\nonumber {R_{h(0)}}_\ast(X_1 \big|_{\gamma_1(0)})[f] &=
X_1\big|_{\gamma_1(0)} [f \circ R_{h(0)}] \\
\nonumber &= \dd{t} f \circ R_{h(0)} \circ \gamma_1(t) \Big|_{t=0} \\
\nonumber &= \dd{t}f(\gamma_1(t)h(0))\Big|_{t=0},
\end{align}
from which we deduce
\begin{equation}
\label{eqn:firstterm}{R_{h(0)}}_\ast(X_1 \big|_{\gamma_1(0)}) =
\dd{t}(\gamma_1(t)h(0))\Big|_{t=0}.
\end{equation}
Also we have
\begin{align}
\nonumber (h(0)^{-1}\extd h
(X\big|_{\gamma(0)}))^\sharp\big|_{\gamma_2(0)} &= \dd{t}
  \big(\gamma_2(0) e^{th(0)^{-1}\extd h
  (X|_{\gamma(0)})}\big)\Big|_{t=0} \\
\nonumber &= \gamma_2(0)h(0)^{-1}\extd h (X\big|_{\gamma(0)}) \\
\label{eqn:secondterm}&= \gamma_2(0)h(0)^{-1}\dd{t}h(t)\Big|_{t=0}.
\end{align}
Thus if we use both \refec{eqn:firstterm}{eqn:secondterm} in
\refe{eqn:bothterms} we find
\begin{displaymath}
X_2\big|_{\gamma_2(0)} = {R_{h(0)}}_\ast(X_1 \big|_{\gamma_1(0)}) +
(h(0)^{-1}\extd h (X\big|_{\gamma(0)}))^\sharp\big|_{\gamma_2(0)}.
\end{displaymath}
Note that the point $t=0$ is not special, and in general we have
\begin{equation}
X_2\big|_{\gamma_2(t)} = {R_{h(t)}}_\ast(X_1 \big|_{\gamma_1(t)}) +
(h(t)^{-1}\extd h (X\big|_{\gamma(t)}))^\sharp\big|_{\gamma_2(t)}.
\end{equation}
\refe{eqn:Xmakehorizontal} follows immediately from this result by
choosing $\gamma_1(t)=L(\gamma(t))$ and $\gamma_2(t)=\tilde{\gamma}(t)$,
whereas \refe{eqn:Xchangesection} follows by choosing
$\gamma_1(t)=L(\gamma(t))$ and $\gamma_2(t)=L'(\gamma(t))$.

\providecommand{\href}[2]{#2}\begingroup\raggedright\endgroup


\begin{thebibliography}{10}

\bibitem{coset}
N.~Alonso-Alberca, E.~Lozano-Tellechea, and T.~Ort\'{\i}n, ``Geometric
  construction of Killing spinors and supersymmetry algebras in homogeneous
  spacetimes,'' {\em Class. Quant. Grav.} {\bf 19} (2002) 6009--6024,
  \href{http://www.arXiv.org/abs/hep-th/0208158}{{\tt hep-th/0208158}}.

\bibitem{hep-th/0503196}
L.~Andrianopoli, S.~Ferrara, M.~A. Lle\'{o}, and O.~Maci\'{a},
``Integration of
  massive states as contractions of non-linear $\sigma$-models,'' {\em J. Math.
  Phys.} {\bf 46} (2005) 072307,
  \href{http://www.arXiv.org/abs/hep-th/0503196}{{\tt hep-th/0503196}}.

\bibitem{bluebook}
I.~L. Buchbinder and S.~M. Kuzenko, {\em Ideas and methods of
supersymmetry and
  supergravity}.
\newblock Institute of Physics Publishing, 1995.

\bibitem{west}
P.~West, {\em Introduction to supersymmetry and supergravity}.
\newblock World Scientific, 1990.

\bibitem{dewitt}
B.~DeWitt, {\em Supermanifolds}.
\newblock Cambridge Monographs on Mathematical Physics. Cambridge University
  Press, 1992.

\bibitem{nakahara}
M.~Nakahara, {\em Geometry, Topology and Physics}.
\newblock Institute of Physics Publishing Bristol and Philadelphia, 2002.

\bibitem{steenrod}
N.~Steenrod, {\em The Topology of Fibre Bundles}.
\newblock Princeton University Press, 1951.

\bibitem{goeckelerschuecker}
M.~G\"ockeler and T.~Sch\"ucker, {\em Differential Geometry, Gauge
Theories,
  and Gravity}.
\newblock Cambridge University Press, 1987.

\bibitem{lightblue}
L.~Castellani, R.~D'Auria, and P.~Fr\'e, {\em Supergravity and
Superstrings, A
  Geometric Perspective}, vol.~1: Mathematical Foundations.
\newblock World Scientific, 1991.

\bibitem{hep-th/0007099}
P.~Claus, J.~Rahmfeld, H.~Robins, J.~Tannenhauser, and Y.~Zunger,
``Isometries
  in anti-de Sitter and conformal superspaces,'' {\em JHEP} {\bf 0007} (2000)
  047, \href{http://www.arXiv.org/abs/hep-th/0007099}{{\tt hep-th/0007099}}.

\bibitem{hep-th/9912277}
L.~Castellani, ``On $G/H$ geometry and its use in M-theory
compactifications,''
  {\em Annals Phys.} {\bf 287} (2001) 1--13,
  \href{http://www.arXiv.org/abs/hep-th/9912277}{{\tt hep-th/9912277}}.

\bibitem{hep-th/9812087}
P.~Claus and R.~Kallosh, ``Superisometries of the $adS \times S$
superspace,''
  {\em JHEP} {\bf 9903} (1999) 014,
  \href{http://www.arXiv.org/abs/hep-th/9812087}{{\tt hep-th/9812087}}.

\bibitem{sym_coset_space}
L.~Castellani, L.~J. Romans, and N.~P. Warner, ``Symmetries of coset
spaces and
  Kaluza-Klein supergravity,'' {\em Annals Phys.} {\bf 157} (1984) 394--407.

\bibitem{supersphere}
A.~F. Schunck and C.~Wainwright, ``A geometric approach to scalar field
  theories on the supersphere,'' {\em J. Math. Phys.} {\bf 46} (2005) 033511,
  \href{http://www.arXiv.org/abs/hep-th/0409257}{{\tt hep-th/0409257}}.

\bibitem{thepaper}
G.~Landi, ``Projective modules of finite type over the supersphere
$S^{2,2}$,''
  {\em Diff. Geom. Appl.} {\bf 14} (2001) 95--111,
  \href{http://www.arXiv.org/abs/math-ph/9907020}{{\tt math-ph/9907020}}.

\end{thebibliography}
\end{document}